\documentclass[preprint]{aastex}

\slugcomment{\today}

\newcommand{\eg}{\mbox{e.g.}}
\newcommand{\ie}{\mbox{i.e.}}

\newcommand{\skipthis}[1]{}

\def\micron{\hbox{$\mu$m}}
\newcommand{\be}{\begin{equation}}
\newcommand{\ee}{\end{equation}}
\newcommand{\e}{et al.\ }

\def\ie{{i.e.\ }}

\usepackage{epsf}
\usepackage{textcomp}

\begin{document}
\shorttitle{NIR Variable YSOs in Cygnus OB7}

\title{Near-infrared Variability among YSOs in the 
Star Formation Region Cygnus OB7}

\author{Scott J. Wolk}
\affil{Harvard--Smithsonian Center for Astrophysics, 60 Garden
Street, Cambridge, MA 02138}

\author{Thomas S. Rice}
\affil{Harvard--Smithsonian Center for Astrophysics, 60 Garden
Street, Cambridge, MA 02138}

\author{Colin Aspin}
\affil{Institute for Astronomy, University of Hawaii at Manoa, 640 N Aohoku Pl, Hilo, HI
96720}

\begin{abstract}

  We present an analysis of near-infrared time-series photometry in $J$, $H$, and $K$ bands  for about 100 epochs of a $1^\circ \times 1^\circ$ region of the Lynds 1003/1004 dark cloud in the Cygnus OB7 region.  Augmented by data from the  Wide-field Infrared Survey Explorer (WISE),  we identify 96 candidate disk bearing young stellar objects (YSOs) in the region.  Of these, 30 are clearly Class I or earlier.     Using the Wide-Field imaging CAMera (WFCAM) on the United Kingdom InfraRed Telescope (UKIRT), we were able to obtain photometry over three observing seasons, with photometric uncertainty better than 0.05 mag down to $J \approx 17$.  
We study detailed light curves and color trajectories of $\sim$50 of the YSOs in the monitored field.
We investigate the variability and periodicity of the YSOs and find the data are consistent with all YSOs being variable in these wavelengths on time scales of a few years.   We
divide the variability into four observational classes: 
1) stars with periodic variability stable over long timescales, 
2) variables which exhibit short-lived cyclic behavior, 
3) long duration variables, and 
4) stochastic variables. Some YSO variability defies simple classification. 
We can explain much of the observed variability as being due to dynamic and rotational changes in the disk, 
including an asymmetric or changing blocking fraction,  changes to the inner disk hole size, as well as changes to the accretion rate. 
Overall, we find that the  Class~I:Class~II ratio of the cluster is consistent with an age of $<$ 1~Myr, with at least one individual, wildly varying, source $\sim 100,000$ yr old. We have also discovered  a Class~II eclipsing binary system with a period of 17.87 days.  
\end{abstract}



\section{Introduction}
\label{sec:Intro}
Young stellar objects (YSOs) were first highlighted because they are irregular optical variables \citep{Joy42, Joy45}. As our understanding of these objects has grown we now consider them as composite systems.  In addition to a star, there may be an accretion layer, a disk, and an outer envelope.   Strong magnetic fields in the star can induce cool starspots, while the accretion can create hot spots in the photosphere. These individual components are dynamic, rotate and experience various wave phenomena.  Changes in any component may induce changes in luminosity. From optical studies, \citet{Her94} identified three types of variability specifically associated with the most well known YSO class -- T Tauri Stars.  One type of variability (called by \citet{Her94} ``Type I'')  is characterized by a low level periodic modulation of the stellar flux (a few 0.1 mag) and results from the rotation of a cool spotted photosphere. A second type of variability is associated with short-lived accretion-related hot spots at the stellar surface of TTs;  this ``Type II'' variability has larger photometric amplitudes (up to 2 or 3 mag\ in extreme cases) and is  most often irregular.  They also noted some stars 
 with month--long dips, presumed to be due to disk eclipses,  which they labeled ``Type III''.

 Most variability studies have used optical monitoring.  Type I variability allows direct measurement of the rotation period of YSOs by optical monitoring of the stellar flux. It has been in use regularly for this purpose over the last two decades \citep[e.g.][]{Att92, Reb01, Coh04, Par09}.    However, the nature of YSOs favor near and mid-IR wavelengths. 
 Near-infrared (NIR) studies of young stars allow for the direct detection of optically thick protoplanetary disks around these stars via excess$K$-band flux \citep{Lad92}. 
The study of near-infrared variability in young stars allows us to study changes in those disk structures. Studies of the Orion A molecular cloud and the Chameleon I molecular cloud 
established that NIR variability is present in YSOs \citep{Car01, Car02}. 
 In Orion, as many as 93\% of the variable stars were identified to be young stars, and a strong connection was established between variability and near-infrared excess. 
Studies of individual YSOs such as AA Tau and its analogs reveal insights into magnetospheric accretion processes linked to inner-disk dynamics 
\citep{Bou03, Bou07, 2010MNRAS.409.1347D}.
Other types of stars such as EX Lup \citep{2010ApJ...719L..50A} and V1118 Ori \citep{2010A&A...511A..63A}  that exhibit large, eruptive mass-accretion, due to infall events of $M > 0.1 M_\oplus$  are easily studied in the near-infrared. During outburst, their near-infrared emission is dominated by hotspot radiation,  emission reprocessed in the disk. Further, the inner disk edge appears to move inwards and brightens in the near-infrared.

Recent mid-IR variability surveys with $Spitzer$ 
 provide insights into physical processes of young stars over short ($\sim$40 day) timescales. These surveys find 60--70\% of YSOs with infrared excess are variable \citep{Mor09, Mor11, Fla12}. Importantly, \citet{Mor11} also find a number of dust-eclipse events suggestive of AA Tau, giving insight into the structure and behavior of protoplanetary disks around stars of this age, as well as the importance of magnetically driven accretion onto young stars.   \citet{Sch12} studied the NIR variability of several young clusters including the ONC, NGC 1333, IC 348 and $\sigma$ Orionis. He finds variability amplitudes are largest in NGC 1333, presumably because it has the youngest sample of YSOs. The frequency of highly variable objects also increases with the time window of the observations.

We have carried out a near-IR JHK survey of the photometric variability of objects in the Lynds~1003/1004 dark cloud within Cygnus OB7 (Cyg~OB7) which we monitored for nearly two years.
Cyg~OB7,  at a distance of around 800 pc \citep[distance modulus $\mu$ = 9.5][]{Asp09}, is the nearest of nine OB associations identified in Cygnus. Cyg~OB7 contains the dark clouds Lynds 1003 and Lynds~1004; previous studies of the cluster in this dark cloud have confirmed it as a region of active star formation. Thirteen T Tauri stars have been identified via near-infrared spectra \citep{Asp09}, 
two outbursting FUor objects 
\citep{1997AJ....114.2700R, 2006A&A...455.1001M, 2008AJ....135.1421G}, 
protostellar millimeter-emitting cores \citep{2011AJ....141..139A},
 as well as Herbig-Haro outflows \citep{2003A&A...412..147M, 2010AJ....139..969M} are also present. 

In a recent paper, Rice \e (2012; hereafter Paper~I) presented first results of this multi-epoch monitoring campaign of the Cyg OB7 region.
We examined photometry of 9,200 stars, across 120 nights with precision better than 4\%, identifying a sample of 30 YSOs, including 6 from \citet{Asp09}.  We found 93\% of the YSOs varied significantly.    The variability observed included color changes which caused sources to transit the disk/photosphere demarkation in the IR color-color diagram.  While some of the changes may be rotationally modulated surface spots (hot or cold), other color changes were identified with changes in the disk structure, others required the inclusion of additional  reddening as well, and still others appeared still more complicated.    

In a companion paper, Wolk \e (2013; hereafter Paper~II) analyzed the subset of variable stars in this field that lack evidence of disks. Using the long-baseline, moderate-cadence time-series photometry, we identify periodic and stochastic variability in about 150 field stars.   We conclude the rate of field star variability in the near infrared is about 1.6\% in this field. Periodic variability is seen in about one-third to one-half of the variable field stars and is dominated by evolved eclipsing binary systems.

In this paper,  we present a follow-on to the analyses presented in Papers~I \& II.  We expand the $J$, $H$, and $K$ variability study towards Cyg OB7 to include the examination of all disk bearing sources.  We have also used period analysis on the YSOs. With long-baseline, moderate-cadence time-series photometry, we identify young stars with periodic, quasi-periodic and stochastic variability.
We use empirical evidence to suggest that these are caused by rotational, dynamic, and eruptive events.
In \textsection \ref{sec:data} we briefly restate the photometric reliability of the WFCAM data.
  We also bring newly released Wide Field Infrared Survey Explorer  (WISE; Wright et al. 2010)  data to bear in identifying additional YSOs in the field. 
We then move on to discuss variability among the YSOs. 
Repeating the techniques used in Papers I \& II, we find that the NIR variability among YSOs can be described by a limited number of families.  These families include periodic, quasi-periodic, long duration and stochastic variables. 
Finally in \S 4 we 
discuss the different phenomena and comment on possible physical mechanisms.

 
\section{Observation and Data Reduction}
\label{sec:data}

The dataset in use here was fully described in Paper~I.  In brief,  $J$, $H$, and $K$ observations of the Cygnus OB7 region
 were obtained using the Wide Field Camera (WFCAM) instrument on the United Kingdom InfraRed Telescope (UKIRT), an infrared-optimized 3.8 meter telescope atop Mauna Kea, Hawaii at 4,200 meters elevation. The field of view of the study is about a 1 degree square centered on 21h00m +52$^{\rm o}$30\arcmin\ (J2000.) near the ``Braid Nebula Star''. Our data consist of WFCAM observations taken from May 2008 to October 2009 in three observing seasons as part of a special observation program. Data were taken on 124 nights; of these, 100 were deemed to be of high-quality in that they were internally consistent in color and magnitude to within 1\% against night--to--night variations.  Using the three seasons of UKIRT observations we obtained high-quality photometry -- defined as 
having errors less than 5\% at $J \sim 17$, $H \sim 17$, and $K \sim 16.5$ and no error flags --
on 9,200 stars. In Paper~I, we did not include stars fainter than $J=17$ nor brighter than $J =11$ as bright stars would saturate in many epochs if the conditions were especially good seeing or if the star brightened. 

In Paper~I, YSOs were identified using the classical $J-H$, $H-K$ color magnitude diagram (Lada \& Adams 1992) which identifies stars with optically thick disks at $K$-band.  A total of 30 YSOs were identified in this way including seven which transited the photosphere/disk boundary as a result of observed variability (Rice \e 2012).   Variability was quantified for all sources in the field using the Stetson index (Stetson 1996).\footnote{We label this index as $S$.  Stetson originally used $J$, but this can be confused with the photometric filter (\eg\ Carpenter \e 2001)}   This is a method for quantifying correlated variability within samples which includes multiple colors each with different error characteristics.  The resultant value is zero for a constant source and exceeds 1 for a source with strong, correlated variability.\footnote{\citet{Car01} compared the Stetson index to $\chi^2$ fits and concluded that $S>0.55$ was sufficient to confirm variability.  In Paper~I we concluded that $S>1.0$ was preferable because the more conservative value prevented false detections in large samples. In the present paper we identify 3 objects with 0.55 $< S < $1.0.} 
 In Paper~I we indicated about 160 variable stars including 28 of the YSOs. 

As described in Paper~II we search for periodicity among all variable sources using two techniques: the Lomb-normalized periodogram \citep[LNP; e.g.][]{1989ApJ...338..277P} and the Fast $\chi^2$  algorithm \citep[F$\chi^2$;][]{2009ApJ...695..496P}.  Both are suited to period analysis on unevenly sampled data such as ours. The LNP method is a useful and popular way to analyze periodicity with an easily interpreted periodogram, which identifies multiple candidate periods and their relative probability. This tended to be more reliable on longer period data (period $>1$ day). The F$\chi^2$ method was empirically more reliable for complex variables such pulsating stars and for short-period ($\lesssim$ 1 day), highly stable,  non-sinusoidal variables, such as eclipsing binaries.

\subsection{WISE Data}

Our goals for this paper include verification of the Class II nature of the sources detected in Paper~I and investigation
of variability among other disked stars in the field.  Paper~I was limited to YSOs in the fairly limited magnitude range discussed in the previous subsection, and with high quality data in all three NIR bands. Recently the WISE point source catalog became available \citep{Cut12a}.
WISE conducted a survey of the entire sky from  7 January 2010 to 6 August 2010, in the 3.4, 4.6, 12 and 22 \micron\  bandpasses (hereafter $W1$, $W2$, $W3$ and $W4$). WISE achieved 5$\sigma$ point source sensitivities better than 16.6, 15.6, 11.3 and 8.0 ($W1$ -- $W4$ respectively; Vega mag)  in unconfused regions with an angular resolution of 6.1\arcsec, 6.4\arcsec, 6.5\arcsec and 12.0\arcsec\ in the four bands. The WISE Source Catalog (WSC) contains the attributes for over $5\times 10^8$ 
point-like and resolved objects. Catalog sources are required to have a measured SNR$>5$ in at least one band, and to meet other criteria to insure a high degree of reliability \cite{Cut12b}.

We extracted about 35,400 sources from the WISE Source Catalog in a 1.42 degree diameter centered at RA 315.1 Dec 52.5  which includes the entire square degree survey area and 0.62 square degrees outside the RWA survey area.  
Table~\ref{Table:WISE:RWA} provides the mid-IR colors of the YSOs identified in Paper~I (``RWA~1--30"), while Table~\ref{Table:known}  lists near and mid IR colors for nine other previously known YSOs in the field.  
Errors are generally quite small, exceeding 3\% in $W1$ and $W2$ only for RWA 8, 14 and 28 with a largest error of 14\%.
Errors in the redder channels are generally $<5\%$ and only exceed 20\% for RWA 8, 10, 14, 21, 23 and 33.  By far the two largest errors occur on RWA10 and RWA 14 which have errors of about 50\% in $W3$ and $W4$ respectively. 
RWA~9 is surprisingly excluded from the WISE PSC although the source is present in WISE atlas images.  



\begin{deluxetable}{lllllcccll}

  \tabletypesize{\scriptsize}
  
  \tablecaption{WISE colors of RWA stars\label{Table:WISE:RWA}}
  \tablewidth{0pt}

  \tablehead{
      \colhead{RWA} &
   \multicolumn{4}{c}{Photometry} &
     \multicolumn{3}{c}{Colors} &
      \colhead{Other}  &
    \colhead{Class}\\
 
    \colhead{\#} &
    \colhead{$W1$ } &
    \colhead{$W2$ } &
    \colhead{$W3$ } &
    \colhead{$W4$ } &
    \colhead{$W1-W2$} &
    \colhead{$W2-W3$ } &
    \colhead{$W2-W4$ } &
\colhead{Name      }  & 
    \colhead{~ }
  }
\startdata
1 & 9.83 & 9.22 & 7.10 & 4.48 & 0.61 & 2.12 & 4.74 &CN 2\tablenotemark{a} & Class IIR \\ 
2 & 9.19 & 7.23 & 3.65 & 1.06 & 1.96 & 3.58 & 6.16 & 21005+5217\tablenotemark{b}  & Class I \\ 
3 & 11.76 & 11.16 & 8.86$^*$ & 5.56 & 0.61 & 2.30 & 5.59 & ~ & Class IIR \\ 
4 & 10.26 & 9.67 & 7.24 & 4.59 & 0.59 & 2.43 & 5.08 & ~ & Class IIR \\ 
5 & 9.92 & 9.21 & 7.06 & 4.32 & 0.71 & 2.14 & 4.88 & CN 3S\tablenotemark{a}  & Class IIR \\ 
6 & 12.03 & 11.55 & 9.70$^*$ & 7.65$^*$ & 0.48 & 1.85 & 3.91 & ~ & Class II \\ 
7 & 11.17 & 10.03 & 7.06 & 3.91 & 1.13 & 2.98 & 6.12 & 2100019+523515\tablenotemark{c}  & Class I \\ 
8 & 13.48 & 12.97$^*$ & 10.59$^*$ & 7.85$^*$ & 0.51 & 2.38 & 5.12 & ~ & Class IIR \\ 
9 & \ldots & \ldots & \ldots & \ldots & \ldots & \ldots & \ldots & \ldots \\ 
10 & 12.59 & 12.15 & 10.38$^*$ & 7.48 & 0.44 & 1.77 & 4.67 & ~ & Class II \\ 
11 & 9.75 & 8.65 & 5.74 & 3.17 & 1.10 & 2.91 & 5.48 & ~ & Class I \\ 
12 & 9.29 & 7.98 & 4.98 & 2.40 & 1.31 & 3.00 & 5.58 & 2059518+524020\tablenotemark{c} & Class I \\ 
13 & 9.88 & 9.33 & 7.42 & 5.46 & 0.54 & 1.91 & 3.88 &CN 7\tablenotemark{a} & Class II \\ 
14 & 13.56$^*$ & 13.21$^*$ & 10.86$^*$ & 8.46$^*$ & 0.35 & 2.35 & 4.75 & ~ & Class IIR \\ 
15 & 10.75 & 9.32 & 6.54 & 3.48 & 1.44 & 2.78 & 5.84 &  CN 1\tablenotemark{a}  & Class I \\ 
16 & 10.08 & 9.63 & 8.09 & 6.13 & 0.46 & 1.54 & 3.50 & ~ & Class II \\ 
17 & 9.97 & 8.56 & 5.51 & 2.79 & 1.40 & 3.05 & 5.78 & 2058500+523255\tablenotemark{c}  & Class I \\ 
18 & 10.16 & 9.64 & 8.25 & 6.18 & 0.52 & 1.39 & 3.47 & ~ & Class II \\ 
19 & 9.21 & 7.94 & 5.53 & 3.12 & 1.27 & 2.40 & 4.82 & CN 6\tablenotemark{a} & Class I \\ 
20 & 8.46 & 7.70 & 5.28 & 1.75 & 0.76 & 2.42 & 5.96 & CN 8\tablenotemark{a} & Class IIR \\ 
21 & 10.35 & 9.92 & 8.32 & 5.82 & 0.42 & 1.61 & 4.11 & ~ & Class II \\ 
22 & 10.29 & 9.56 & 7.80 & 5.12 & 0.74 & 1.76 & 4.44 & ~ & Class II \\ 
23 & 12.54 & 12.09 & 10.44 & 7.72 & 0.46 & 1.65 & 4.37 & ~ & Class II \\ 
24 & 10.23 & 9.35 & 7.23 & 4.75 & 0.88 & 2.12 & 4.60 & ~ & Class IIR \\ 
25 & 9.93 & 9.53 & 7.94 & 5.90 & 0.40 & 1.59 & 3.63 & ~ & Class II \\ 
26 & 12.46 & 11.46 & 9.16 & 6.37 & 1.00 & 2.30 & 5.09 & ~ & Class I \\ 
27 & 12.30 & 11.78 & 9.87 & 7.09$^*$ & 0.52 & 1.91 & 4.69 & ~ & Class II \\ 
28 & 14.24$^*$  & 13.80$^*$  & 9.80& 7.31$^*$ & 0.44 & 4.00 & 6.49 & ~ & Class IIR \\ 
29 & 12.52 & 12.17 & 10.37$^*$ & 7.70$^*$ & 0.35 & 1.80 & 4.47 & ~ & Class II \\ 
30 & 12.31 & 11.50 & 9.34 & 6.89$^*$ & 0.81 & 2.16 & 4.61 & ~ & Class IIR \\ 
\enddata
  \tablenotetext{*}{Photometric errors are larger than the typical photometric errors of $< 3\%$ for $W1$ and $W2$  or $<20\%$ for 
  $W3$ and $W4$.}
    \tablenotetext{a}{From Aspin et al.\ 2009.}
   \tablenotetext{b}{AKARI ID}
   \tablenotetext{c}{IRAS ID}
 \end{deluxetable}



\begin{deluxetable}{lllllllllll}

  \tabletypesize{\scriptsize}
  
  \tablecaption{2MASS and WISE colors of other YSOs in the monitored field.\label{Table:known}}
  \tablewidth{0pt}

  \tablehead{
   \colhead{Object ID} &
    \colhead{$J$ } &
    \colhead{$H$ } &
    \colhead{$K$ } &
    \colhead{$W1$ } &
    \colhead{$W2$ } &
    \colhead{$W3$ } &
    \colhead{$W4$ } &
    \colhead{Other Name} &
    \colhead{Class}  \\
       \colhead{2MASS (J2000.)} &
    \colhead{(mag)} &
    \colhead{(mag)} &
    \colhead{(mag)} &
    \colhead{(mag)} &
    \colhead{(mag)} &
    \colhead{(mag)} &
    \colhead{(mag)} &
    \colhead{} &
    \colhead{} 
  }
  \startdata
205821.09+522927.7 &  11.54 & 9.81 & 8.31 & 5.46 & 3.38 & 0.57 & -1.50 & HH381 IRS\tablenotemark{a} & Class I\\
210011.59+521817.3 & 11.38 & 10.48 & 10.02 & 9.27 & 8.79 & 7.31 & 6.04 & Cyg~19\tablenotemark{a} & Class II \\ 
210003.76+523429.0 &  13.32 & 11.28 & 9.99 & 8.36 & 7.52 & 5.31 & 3.17 & CN3N\tablenotemark{a} & Class I\\
210021.13+522705.2 &  12.43 & 12.04 & 11.96 & \ldots  & \ldots  &\ldots  & \ldots & IRAS 15S\tablenotemark{a,c} & \dots\\
210021.40+522709.4 &  11.43 & 10.50 & 9.72 & 7.89 & 5.75 & 3.01 & 0.31 & IRAS 15N\tablenotemark{a,c} & Class I\\
210021.42+522257.1 & 17.36 & 16.69 & 14.05 & 11.27 & 8.70 & 5.30 & 3.20 & I20588+5221\tablenotemark{b}  & Class I \\ 
210025.25+523017.0 & \ldots & \ldots & \ldots & 9.55 & 7.26 & 4.12 & 1.14 & Braid Star\tablenotemark{a} & Class I \\ 
210038.77+522757.5 & \ldots & \ldots & \ldots & 11.89 & 8.89 & 6.58 & 3.52 & core\#33JI\tablenotemark{b} & Class I \\ 
210042.38+522600.8 & 18.26 & 15.51 & 14.06 & 11.82 & 9.63 & 7.24 & 3.94 & IRAS 14\tablenotemark{a} & Class I \\ 
\enddata

  \tablenotetext{a}{\citet{Asp09}}
  \tablenotetext{b}{Khanzadyan et al.\ (2012)}
  \tablenotetext{c}{Sources are not resolved in WSC or 2MASS. WFCAM median colors are used for J,H and K. WISE colors are presumed dominated by the brighter source IRAS~15N.}
  \tablecomments{Typical photometric errors are $< 2\%$  for $W1$ and $W2$ and $<20\%$ for
  $W3$ and $W4$. }

\end{deluxetable}

Previously, with both near and mid-IR surveys, color--color and color--magnitude diagrams have proven to be powerful tools in identifying Class I and Class II objects \citep{Lad92, All04, Meg04, Gut09}. This is because stars without disks have spectral energy distributions that  decrease with decreasing energy.  Stars with modest disks (\ie\ Class II objects) have SEDs which fall less slowly. As the classes work back toward Class 0 the SEDs flatten out and can even rise, indicating the most energy is radiated at the longest wavelengths \citep{Rob06}.  Recently \citet{Koe12} has used multi-color filtering akin to that used for IRAC and MIPS \citep{Gut09} to identify PMS stars in the WISE fields. 
To locate YSO candidates in Cyg~OB7 we followed an approach similar to \cite{Koe12} but with some modifications based on the field.

\begin{figure}
  \includegraphics[width=6.5in]{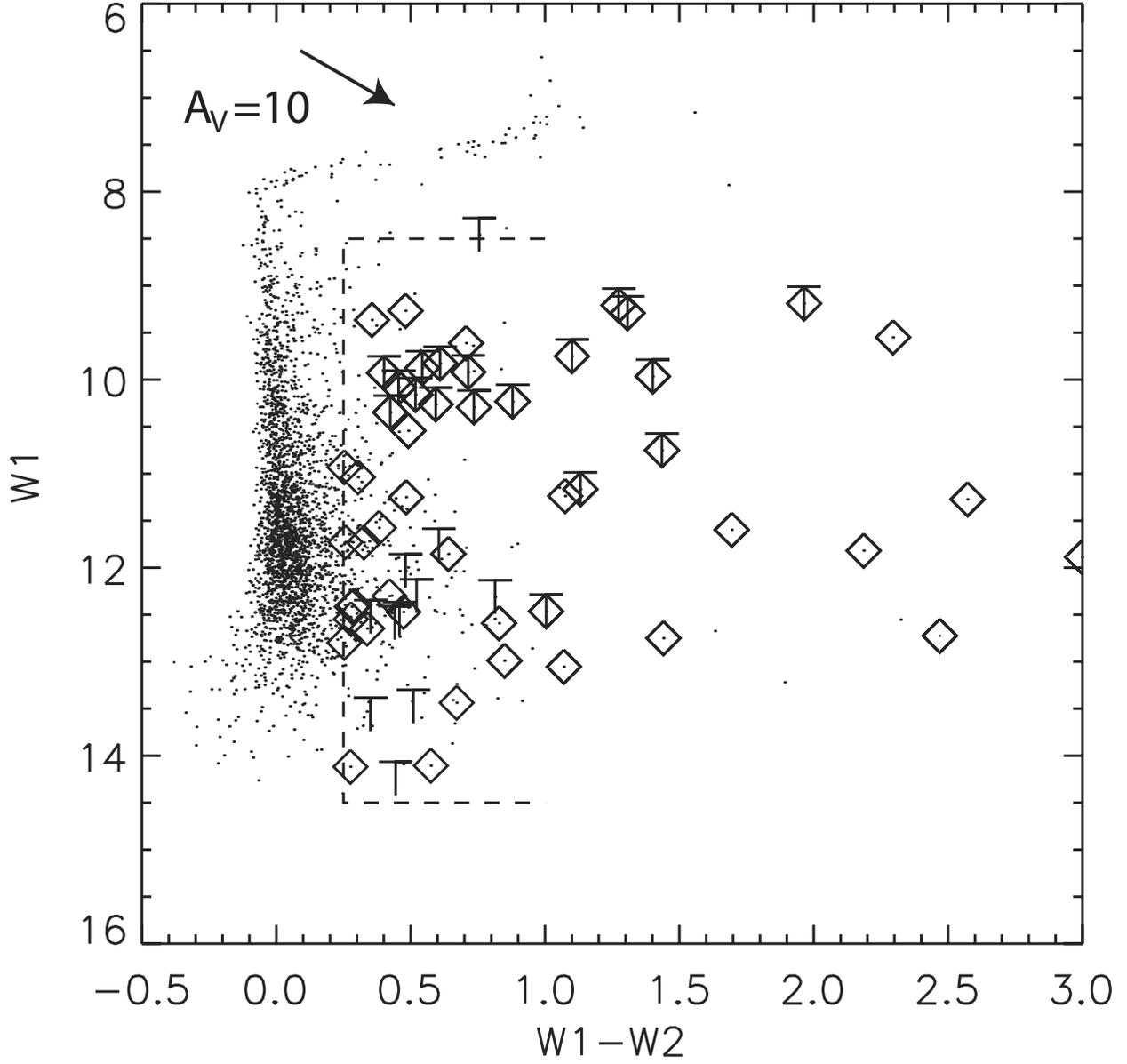}
  \caption{
Mid--IR color-magnitude diagram for 2,769 bright WISE detected sources within the $1^{\rm o} \times  1^{\rm o}$ RWA  field. 
'T' indicates a star from Paper~I.  Diamonds are candidate YSOs with $W1-W2 > 0.25$.  
The dashed box indicates the YSO discovery region. Small dots within this region indicate sources without confirmation in $W2-W3$. This is usually due to low photometric precision.  A reddening vector of 10 A$_V$ is indicated.
}
  \label{w1w1w2}
\end{figure}

\begin{figure}
    \includegraphics[width=6.5in]{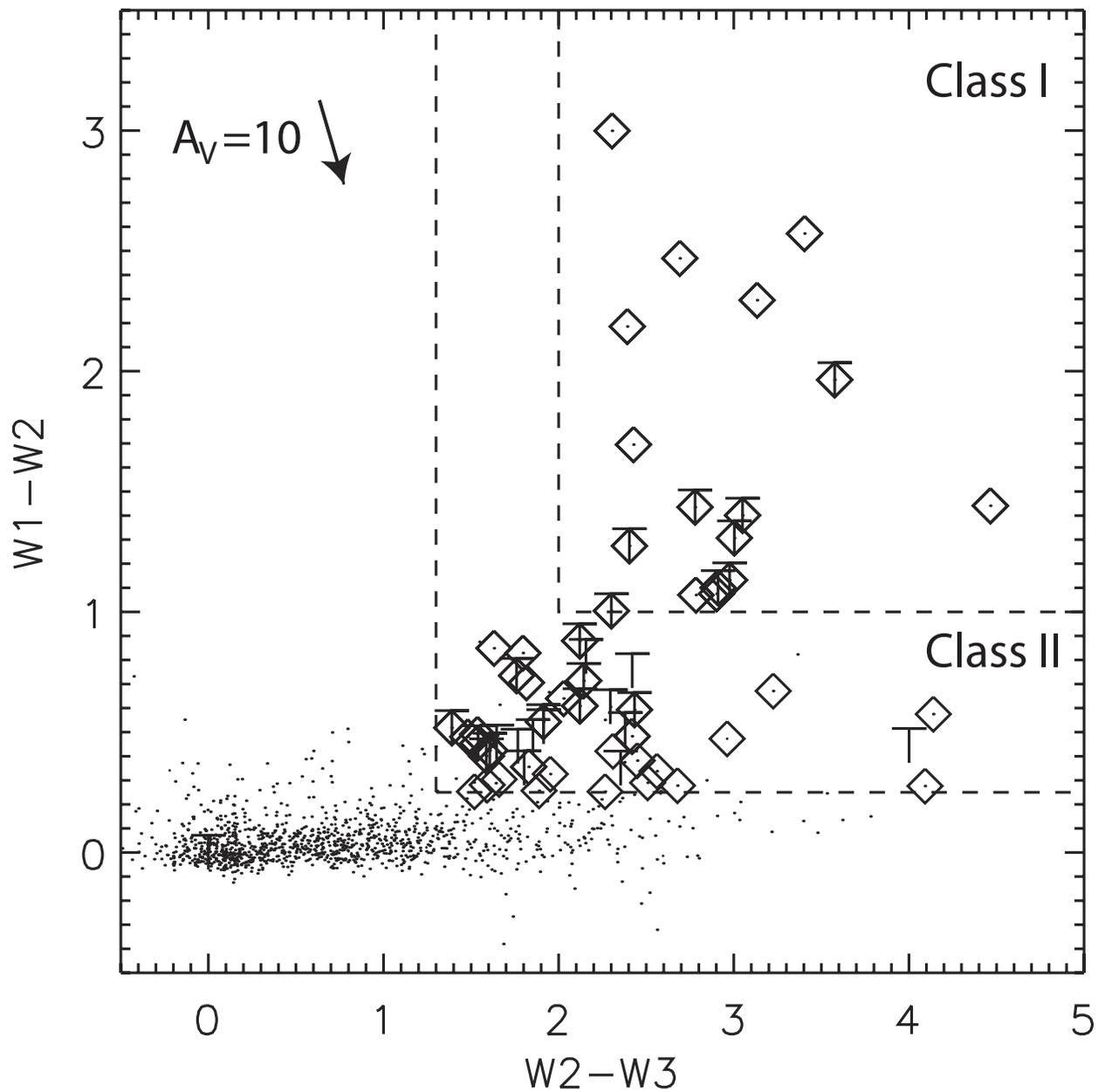}
  \caption{
Mid--IR two-color diagram for WISE sources with good photometry at $W3$ and within the $1^{\rm o} \times  1^{\rm o}$ RWA field. 
Symbols are as in  Fig.~\ref{w1w1w2}.  The dashed lines to the left indicate the Class II region, the dashed lines to the right indicate the Class I/0 region \citep{Koe12}.}
\label{w2w3w1w2}
\end{figure}

To summarize, after excluding faint sources to remove background galaxies, resolved PAH regions and shocks, \citet{Koe12} used a color cut of $W1-W2> 0.25$ and $W2-W3 > 1.0$ to identify YSOs. 
The 29 WISE detected RWA sources all fit within these parameters.
We also filtered out all stars brighter than $W1=8.5$ which tend to have very red colors indicative of a saturation issue.
The known disked objects have $W2-W3>1.3$ and we use {\it this} as the second requirement on YSOs in the region 
(Figs.~\ref{w1w1w2} and \ref{w2w3w1w2}). Further, we also require the errors to be low, $<0.04$ in $W1$ and $< 0.2$ in $W3$. 

In total there are 66 
 strong YSO candidates in the 1 square degree field. 
 \footnote {Relaxing the constraint to $W2-W3>1.0$  added only 3 PMS candidates, all of which were within 3 sigma of the mean $W1-W2$ color.}  
Of these,   39 have been previously identified  \citep{Asp09, Ric12, Kha12}. These are included in Tables 1 and 2. The 27 new YSO candidates in the field studied in Papers~I \& II are listed in Table~\ref{TableWISE2} . 



\begin{deluxetable}{llllllllll}

  \tabletypesize{\scriptsize}
  
  \tablecaption{2MASS and WISE colors of new YSO candidates in the monitored field.\label{TableWISE2}}
  \tablewidth{0pt}

  \tablehead{
   \colhead{Object ID} &
    \colhead{$J$ } &
    \colhead{$H$ } &
    \colhead{$K$ } &
    \colhead{$W1$ } &
    \colhead{$W2$ } &
    \colhead{$W3$ } &
    \colhead{$W4$ } &
    \colhead{Class}  \\
       \colhead{2MASS (J2000.)} &
    \colhead{(mag)} &
    \colhead{(mag)} &
    \colhead{(mag)} &
    \colhead{(mag)} &
    \colhead{(mag)} &
    \colhead{(mag)} &
    \colhead{(mag)} &
    \colhead{} 
  }
  \startdata
205736.61+522117.0 & 12.45 & 11.44 & 10.76 & 9.36 & 9.01 & 7.18 & 5.14 & Class II \\ 
205745.36+521331.4 & 13.19 & 12.58 & 12.30 & 11.73 & 11.47 & 9.58 & 7.20 & Class II \\ 
205811.89+522923.3 & 14.76 & 12.88 & 11.99 & 11.58 & 11.19 & 8.75 & 4.88 & Class IIR \\ 
205816.27+522832.8 & 14.90 & 12.77 & 11.82 & 11.25 & 10.77 & 8.35 & 4.26 & Class IIR \\ 
205837.72+521845.1 & 15.48 & 14.28 & 13.84 & 12.42 & 12.13 & 10.54 & 8.33 & Class II \\ 
205842.10+522545.7 & 17.41 & 15.89 & 14.18 & 11.60 & 9.90 & 7.48 & 4.73 & Class I \\ 
205854.11+522553.3 & 15.44 & 14.00 & 13.31 & 12.80 & 12.55 & 10.28 & 7.08 & Class IIR \\ 
205855.04+523039.6 & \ldots & \ldots & \ldots & 12.73 & 10.26 & 7.56 & 4.00 & Class I \\ 
205904.15+523007.8 & 16.48 & 15.48 & 13.85 & 12.75 & 11.31 & 6.84 & 3.06 & Class I \\ 
205904.29+523302.6 & 16.36 & 14.15 & 12.82 & 12.30 & 11.88 & 9.57 & 7.25 & Class IIR \\ 
205926.05+522607.6 & 15.22 & 13.80 & 13.21 & 12.55 & 12.27 & 9.59 & 5.89 & Class IIR \\ 
205939.84+522003.1 & 15.12 & 13.54 & 12.92 & 12.40 & 12.11 & 9.60 & 6.89 & Class IIR \\ 
210021.39+522625.2 & 16.40 & 15.24 & 14.54 & 13.05 & 11.98 & 9.20 & 4.71 & Class I \\ 
210024.05+521451.0 & 13.99 & 13.04 & 12.62 & 11.85 & 11.21 & 9.19 & 7.16 & Class IIR \\ 
 210025.88+522721.2 & 15.61 & 13.90 & 13.14 & 12.47 & 12.00 & 9.04 & 4.22 & Class IIR \\ 
210031.21+523110.9 & \ldots & \ldots & \ldots & 14.11 & 13.53 & 9.39 & 6.65 & Class IIR \\ 
210031.74+523058.9 & 18.10 & 15.82 & 14.37 & 13.44 & 12.77 & 9.54 & 6.76 & Class IIR \\ 
210035.25+523334.0 & 15.00 & 13.19 & 12.42 & 11.24 & 10.16 & 7.26 & 3.57 & Class I \\ 
210035.82+523348.4 & 14.85 & 13.00 & 12.21 & 11.73 & 11.41 & 9.45 & 4.85 & Class II \\ 
210058.34+522856.1 & 12.33 & 11.08 & 10.48 & 9.61 & 8.91 & 7.09 & 5.30 & Class II \\ 
210110.57+521512.9 & 13.06 & 12.05 & 11.29 & 10.54 & 10.05 & 8.52 & 6.55 & Class II \\ 
210111.73+521509.7 & 13.09 & 12.11 & 11.65 & 11.04 & 10.73 & 9.07 & 6.71 & Class II \\ 
210116.78+522640.4 & 14.31 & 12.25 & 11.33 & 10.93 & 10.68 & 9.16 & 6.65 & Class II \\ 
210233.58+522745.4 & 15.71 & 14.88 & 14.68 & 14.12 & 13.84 & 9.75 & 6.74 & Class IIR \\ 
210246.43+523114.0 & 17.80 & 15.94 & 14.08 & 12.59 & 11.76 & 9.97 & 6.69 & Class II \\ 
210257.89+523344.7 & 15.65 & 13.90 & 13.09 & 12.65 & 12.31 & 9.75 & 7.91 & Class IIR \\ 
210311.54+523124.3 & 17.58 & 15.02 & 13.63 & 12.99 & 12.14 & 10.51 & 6.73 & Class II \\ 
\enddata

  \tablecomments{Typical photometric errors are $< 2\%$  for $W1$ and $W2$ and $<20\%$ for  $W3$ and $W4$. }

\end{deluxetable}

\begin{figure}
\includegraphics[width=6.5in]{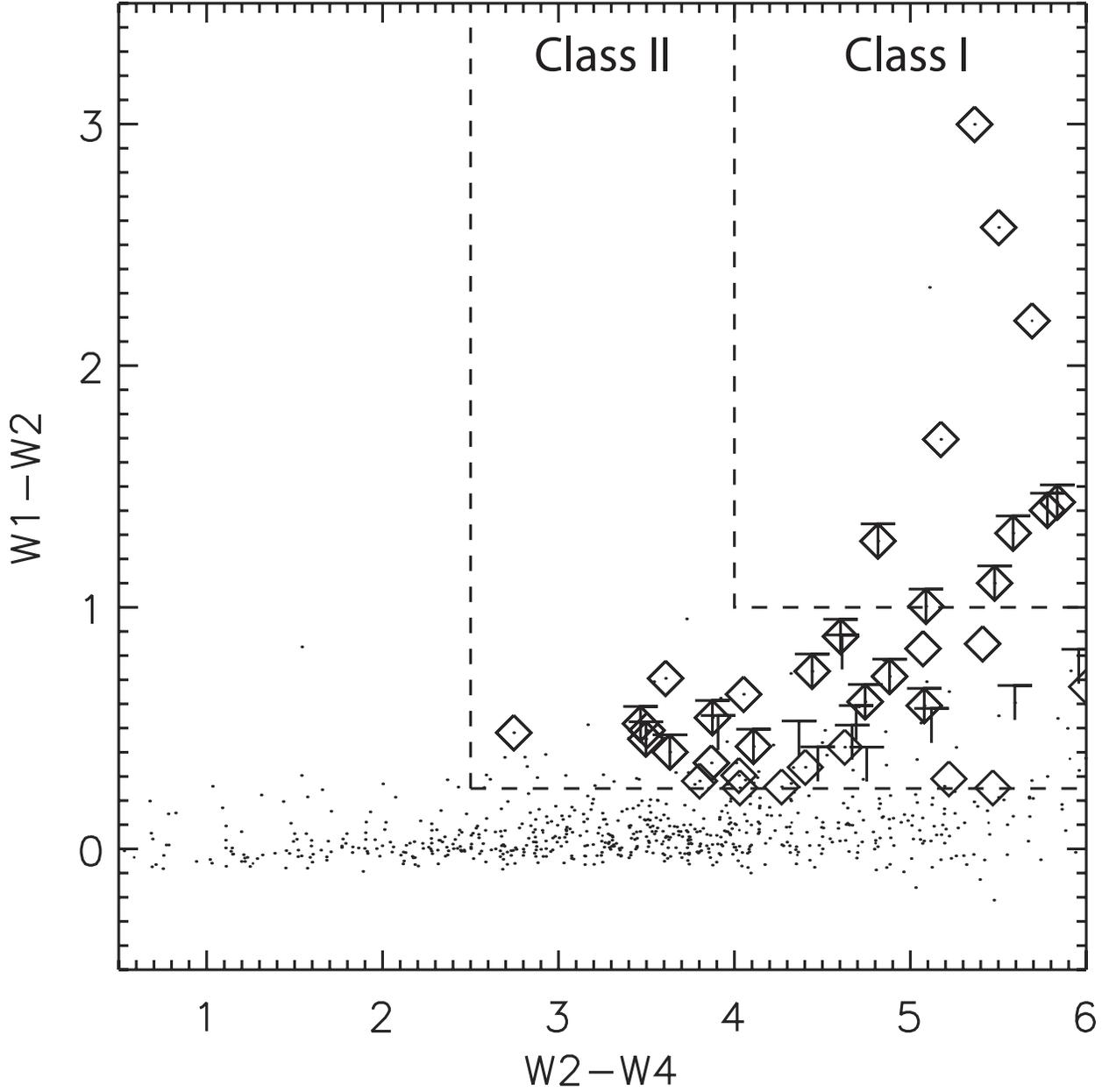}
  \caption{
Mid-IR two-color diagram for WISE sources with good photometry at $W3$ and within .5 degrees of the Braid Nebula star. 
Symbols are as in  Fig.~\ref{w1w1w2}.  The lower dashed lines to the left indicate the Class II region, the upper dashed lines indicate the Class I/0 region.}
  \label{w2w4w1w2}
\end{figure}

\begin{figure}
\includegraphics[width=6.5in]{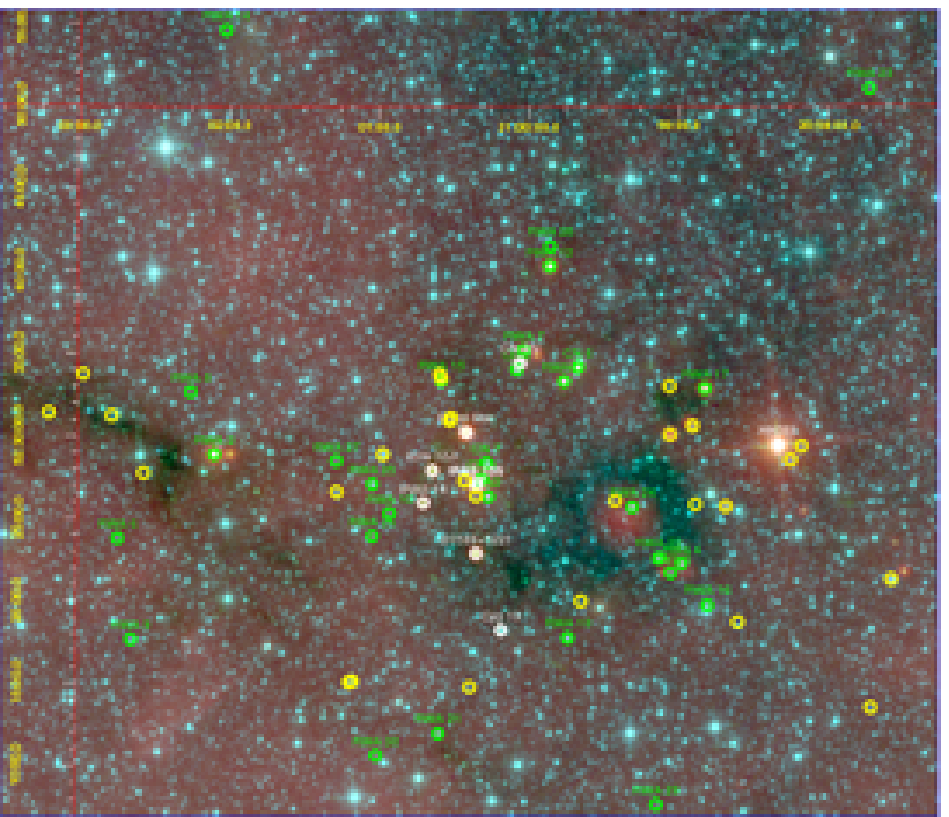}
  \caption{
A WISE atlas image of the Lynds 1003 and 1004 dark clouds covering the field monitored by UKIRT (channels 1, 2 and 4 are blue, green and red respectively).  North is up, East is to the left, L~1004 is the dark region to the east and L~1003 is the dark region to the west. 
Sources from Paper~I are indicated in green, other YSOs from \citet{Asp09}  are indicated in white,  and those detected via the WISE source catalog as described in \S 2.1 are indicated in yellow.
  \label{fig:map}}
\end{figure}

To check the validity of the color cuts, we examined additional WISE colors.  We found that all of our YSO candidates have $W2-W4 >2.5$ and all but one have  $W2-W4 >3.2$   (Fig~\ref{w2w4w1w2}). Likewise eight of the 13 confirmed PMS stars from \citet{Asp09}, including the Braid Nebula star (V2495~Cyg), were recovered.   On the other hand,  three PMS candidates studied and rejected by \citet{Asp09}, CN1S, CN6N and HH627 star,  were not identified as PMS candidates by the WISE color method. Examination of the the surface distribution of the new YSO candidates shows that all the new YSO candidates  lie near the Lynds~1003/1004 dark cloud.  All are within the boarders of the known YSOs, except for a few, just to the east of RWA 2, which are very close to the middle of the L1004 dark cloud (Figure~\ref{fig:map}).

Since the only criteria used in the selection of these stars were based on the WISE point source catalog,  we were able to expand the study region to include all stars in the 1.4 degree diameter circle containing all of the Lynds~1003/1004 dark clouds. We identified 31 additional YSO candidates with $W1-W2 > 0.25$ and  $W2-W3 > 1.3$,  these are listed in Table~\ref{TableWISE3}. While these sources cannot be investigated further with regard to their variability properties, they provide additional insight into the structure of the cluster as a whole (\S 4.4). 
 


\begin{deluxetable}{lllllllllll}

  \tabletypesize{\scriptsize}
  
  \tablecaption{2MASS and WISE colors of new YSOs outside of the monitored field\label{TableWISE3}.}
  \tablewidth{0pt}

  \tablehead{
   \colhead{Object ID} &
    \colhead{$J$ } &
    \colhead{$H$ } &
    \colhead{$K$ } &
    \colhead{$W1$ } &
    \colhead{$W2$ } &
    \colhead{$W3$ } &
    \colhead{$W4$ } &
    \colhead{Notes\tablenotemark{a}} &
    \colhead{Class} \\
       \colhead{2MASS (J2000.)} &
    \colhead{(mag)} &
    \colhead{(mag)} &
    \colhead{(mag)} &
    \colhead{(mag)} &
    \colhead{(mag)} &
    \colhead{(mag)} &
    \colhead{(mag)} &
    \colhead{} &
    \colhead{} 
  }
  \startdata
205558.67+521844.1 & 17.06 & 15.38 & 13.90 & 11.71 & 10.22 & 7.51 & 5.20 &  & Class I \\ 
205621.61+521711.0 & 18.20 & 15.46 & 13.60 & 11.84 & 10.75 & 8.81 & 6.10 & & Class II \\ 
205630.68+521009.3 & 11.04 & 10.70 & 10.53 & 10.08 & 9.64 & 7.27 & 2.99 &  & Class IIR \\ 
205631.44+521028.9 & 13.77 & 13.13 & 12.99 & 11.68 & 10.90 & 7.85 & 3.53 &  & Class IIR \\ 
205633.29+525457.5 & 15.38 & 14.77 & 14.43 & 13.19 & 12.78 & 9.26 & 6.54 &  & Class IIR \\ 
205633.49+523448.7 & 15.69 & 14.26 & 13.65 & 12.76 & 12.40 & 10.76 & 7.97 &  & Class II \\ 
205636.75+523402.3 & 11.59 & 11.02 & 10.56 & 10.03 & 9.44 & 7.42 & 5.36 &  & Class IIR \\ 
205649.26+520333.6 & 15.34 & 14.09 & 13.62 & 12.25 & 11.65 & 9.74 & 6.89 &  & Class II \\ 
205651.59+522041.7 & 18.12 & 15.52 & 14.29 & 11.89 & 10.39 & 7.67 & 4.06 & Western Clump & Class I \\ 
205651.90+522030.8 & 18.20 & 15.48 & 13.50 & 11.34 & 10.07 & 7.38 & 3.30 & Western Clump & Class I \\ 
205651.92+522014.6 & 13.57 & 11.34 & 10.31 & 9.81 & 9.30 & 6.88 & 2.17 & Western Clump & Class IIR \\ 
205652.68+522003.3 & 16.52 & 14.87 & 14.17 & 10.82 & 8.95 & 5.66 & 1.82 & Western Clump & Class I \\ 
205655.33+522022.0 & 16.73 & 14.35 & 12.70 & 11.23 & 10.09 & 8.32 & 3.88 & Western Clump & Class II \\ 
205658.12+522053.8 & 16.64 & 15.06 & 13.51 & 11.64 & 10.51 & 8.64 & 5.79 & Western Clump & Class II \\ 
205700.80+522555.1 & 12.51 & 11.62 & 11.23 & 10.13 & 9.64 & 8.20 & 6.43 &  & Class II \\ 
205701.84+522805.5 & 14.71 & 13.69 & 12.51 & 10.79 & 9.72 & 7.44 & 6.14 &  & Class II \\ 
205920.95+530249.6 & 14.17 & 13.45 & 13.29 & 12.07 & 11.73 & 10.41 & 8.09 &  & Class II \\ 
210316.46+523349.5 & 16.19 & 13.99 & 13.24 & 12.43 & 12.15 & 9.07 & 6.78 &  & Class IIR \\ 
210348.30+521734.3 & 14.62 & 13.20 & 12.72 & 11.56 & 10.91 & 6.58 & 3.83 & Southclump & Class IIR \\ 
210348.45+521745.8 & 15.62 & 14.17 & 13.65 & 11.30 & 10.67 & 6.15 & 3.64 & Southclump & Class IIR \\ 
210349.33+521737.8 & \ldots & \ldots & \ldots & 11.61 & 10.48 & 6.41 & 3.75 & Southclump & Class I \\ 
210359.46+523446.1 & 16.96 & 15.03 & 12.79 & 11.34 & 9.35 & 6.69 & 3.72 & RWA2 tip & Class I \\ 
210401.67+523455.2 & 17.06 & 13.93 & 11.50 & 9.92 & 8.52 & 6.87 & 3.44 & RWA2 tip & Class II \\ 
210407.19+523350.1 & 18.30 & 15.78 & 13.85 & 10.12 & 7.64 & 5.30 & 1.95 & RWA2 tip & Class I \\ 
210408.61+523401.7 & 18.30 & 15.92 & 13.66 & 11.10 & 9.11 & 6.90 & 2.64 & RWA2 tip & Class I \\ 
210411.81+524418.0 & 11.47 & 10.46 & 9.61 & 8.84 & 8.29 & 4.98 & 2.38 &  & Class IIR \\ 
210413.22+524426.3 & 13.47 & 12.79 & 12.54 & 11.43 & 11.00 & 6.88 & 2.90 &  & Class IIR \\ 
210427.40+522156.9 & \ldots & \ldots & \ldots & 9.97 & 8.90 & 5.77 & 2.92 &  & Class I \\ 
210427.91+522203.1 & 12.16 & 11.10 & 10.42 & 9.28 & 8.23 & 5.21 & 2.80 &  & Class I \\ 
210436.23+523447.4 & 15.17 & 13.59 & 13.01 & 12.82 & 12.52 & 10.71 & 6.89 &  & Class II \\ 
\enddata
 \tablenotetext{a}{See \S4.4 for details. }
   \tablecomments{Typical photometric errors are $\sim2\%$ except for
  $W3$ and $W4$ which are $<20\%$ see text for more details. }

\end{deluxetable}

\skipthis{ 
Of the 30 RWA sources, 27 are recovered using the WISE only techniques. The three which were not identified via the WISE color tests as PMS stars all still appear to be YSOs.
For example, RWA~20  is excluded as too bright in W1 although its $W1-W2$ and $W2-W3$ colors (0.76 and 2.42 respectively) are well to the red of our cutoffs. 
RWA~28 is too faint in $W1$ but has $W2-W3$= 4.0.  
}


\citet{Koe12} used data from \citet{Reb10} as a template and found that the reddest parts of the WISE color-color diagrams correspond to protostars \citep[this is also found in \eg][]{Gut09, Rob06}. \citet{Koe12} used color cutoffs of 2.0 and 4.0 in $W2-W3$ and $W2-W4$
respectively to distinguish Class~I and Class~II sources.     All three must agree to identify the object as Class I; if only the redder two colors indicate Class I then the Class is noted as Class IIR. 
Using these color-cuts for 96 YSOs in the WISE field, we find 32 clear Class II sources and 30 clear Class I sources;  most of the remainder show ambiguous colors we identify as Class~IIR in Tables 1--3. 
Restricting ourselves to the 66 YSOs in the UKIRT monitored field, we find 22 clear Class II sources and 20 clear Class I sources. 
The results are consistent and imply the region is quite young -- with nearly equal numbers of stars by Class. 

\section{Results}
\label{sec:results}

In this section, we take a detailed look at the JHK variability of the disked stars in the field with sufficient data quality. This includes the 30 sources from Paper~I,  
and  CN~3N, Cyg~19, HH381~IRS, IRAS~14, IRAS~15N and IRAS~15S \citep{Asp09}. To this we add 13 of the YSOs revealed by analysis of the WISE data (see \S~2.1) which also have good UKIRT monitoring data. Our input catalog is 49 stars, Table~\ref{Table:YSOs} summarizes much of the analysis.     
Columns 2-4  give the full range of variability observed in the $J$, $H$ and $K$ bands and the colors. 
Column 5 is the full range of the color change seen in $J-H$.  
Columns 6--11 list the slope and errors of linear fits to the data in the color--color and color--magnitude plots.   
The slopes are fitted using a Deming regression, which is a special case of a total least squares fit that accounts for the errors on both axis (Deming 1943).   There are some cases in which the data distribution is not sufficiently linear to give a meaningful fit.  In those cases, there is no entry.
Column 12 lists the Stetson index. 
Column 13 lists the period in days or indicates the nature of the variability.  
Nine of thirteen WISE sources and 4/6 \citet{Asp09} sources are found to have a Stetson index $>$ 1.  Combined with the results of paper~I, the overall fraction of sources with strong correlated variability is $84\pm13$\%.



\begin{deluxetable}{lcccccrrrrrrrl}

\tabletypesize{\scriptsize}
  \rotate
  
  \tablecaption{YSOs in the monitored field.\label{Table:YSOs}}
  \tablewidth{0pt}

 \tablehead{
    \colhead{ID} & 
    \colhead{Median $K$ } &
    \colhead{$\Delta J$ } &
    \colhead{$\Delta H$ } &
   \colhead{$\Delta K$ } &
   \colhead{$\Delta (J-H)$ } & 
   \multicolumn{2}{c}{$J/(J-H)$} &
    \multicolumn{2}{c}{$K/(H-K)$} &
     \multicolumn{2}{c}{$(J-H)/(H-K)$} &
    \colhead{$S$}  &
    \colhead{Period}
     \\
    \colhead{~} & 
    \colhead{~} &
    \colhead{(mag.) } &
    \colhead{(mag.) } &
   \colhead{(mag.)} &
   \colhead{(mag.)} & 
   \colhead{slope} &
     \colhead{err.} &
   \colhead{slope} &
       \colhead{err.} &
    \colhead{slope} &
      \colhead{err.} &
   \colhead{~}  &
   \colhead{days/oth.} 
}
\startdata
RWA  Stars  \\
\hline
RWA 1 & 10.98 & 1.85 & 1.46 & 1.13 & 0.45 & 4.97 & 0.13 & 2.81 & 0.16 & 0.87 & 0.03 & 42.96 & 9.11\\
RWA 2 & 11.85 & 2.74 & 2.67 & 1.77 & 0.75 & \ldots & \ldots & \ldots & 0.23 & \ldots & 0.04 & 61.95& LD\\
RWA 3 & 12.21 & 0.11 & 0.14 & 0.09 & 0.12 & 9.62 & 3.14 & \ldots & \ldots  & $-$1.13 & 0.13 & 3.78 &17.87$^a$\\
RWA 4 & 10.82 & 0.34 & 0.34 & 0.34 & 0.12 & 13.08 & 5.54 & \ldots & 1.93 & 0.59 & 0.05 & 8.59 & 6.35\\
RWA 5 & 10.78 & 0.34 & 0.34 & 0.28 & 0.13 & 3.84 & 0.23 & 2.84 & 0.55 & 0.79 & 0.04 & 7.72 & \\
RWA 6 & 12.74 & 0.09 & 0.09 & 0.11 & 0.07 & \ldots & \ldots & $-$3.27 & 0.30 & $-$0.90 & 0.16 & 2.23 & \\
RWA 7 & 12.65 & 0.81 & 0.89 & 0.93 & 0.20 & \ldots & \ldots & \ldots & 2.59 & 0.82 & 0.06 & 30.81& LD\\
RWA 8 & 14.40 & 0.43 & 0.30 & 0.32 & 0.34 & 1.81 & 0.14 & $-$1.16 & 0.13 & 0.74 & 0.10 & 2.77& QP\\
RWA 9 & 12.87 & 0.26 & 0.30 & 0.36 & 0.08 & $-$5.72 & 0.71 & $-$5.58 & 0.45 & 0.69 & 0.09 & 8.65& LD\\
RWA 10 & 13.19 & 0.13 & 0.15 & 0.20 & 0.07 & $-$5.43 & 1.97 & $-$3.27 & 0.39 & 0.62 & 0.10 & 3.33 & \\
RWA 11 & 10.96 & 0.10 & 0.09 & 0.17 & 0.05 & 4.64 & 1.81 & $-$2.28 & 0.29 & $-$0.16 & 0.13 & 3.40 & \\
RWA 12 & 10.74 & 0.73 & 0.61 & 0.57 & 0.15 & 6.08 & 0.82 & \ldots & \ldots  & 0.97 & 0.11 & 14.45&LD/QP\\
RWA 13 & 10.63 & 0.46 & 0.45 & 0.43 & 0.16 & 3.98 & 0.50 & $-$4.50 & 1.08 & 0.78 & 0.06 & 7.95 &9.37\\
RWA 14 & 14.09 & 1.26 & 1.01 & 0.72 & 0.26 & 4.91 & 0.19 & 2.24 & 0.16 & 0.75 & 0.03 & 25.05 & \\
RWA 15 & 12.43 & 1.27 & 0.96 & 0.89 & 0.47 & 3.66 & 0.43 & \ldots & \ldots  & 0.89 & 0.04 & 16.58 & QP\\  
RWA 16 & 11.25 & 0.53 & 0.52 & 0.41 & 0.12 & 5.76 & 0.58 & 6.96 & 2.55 & 0.57 & 0.05 & 11.50 & 4.84\\
RWA 17 & 10.38 & 0.94 & 0.65 & 0.51 & 0.34 & 3.02 & 0.08 & 2.57 & 0.24 & 1.42 & 0.05 & 24.21& LD\\
RWA 18 & 11.12 & 0.66 & 0.57 & 0.61 & 0.18 & 6.81 & 1.08 & \ldots & \ldots  & 0.65 & 0.04 & 11.53 & \\
RWA 19 & 10.66 & 0.61 & 0.75 & 0.58 & 0.25 & $-$3.23 & 0.65 & 4.46 & 0.99 & $-$1.46 & 0.29 & 16.15& LD\\
RWA 20 & 9.79 & 0.61 & 0.58 & 0.61 & 0.13 & 14.05 & 5.79 & $-$7.07 & 1.06 & 0.39 & 0.17 & 14.92 & \\
RWA 21 & 11.16 & 1.20 & 0.95 & 0.67 & 0.28 & 4.33 & 0.19 & 2.63 & 0.24 & 0.90 & 0.03 & 19.91& 3.72\\
RWA 22 & 11.24 & 0.33 & 0.35 & 0.45 & 0.08 & \ldots & \ldots & $-$3.23 & 0.21 & 0.43 & 0.05 & 7.59 & \\
RWA 23 & 13.48 & 0.33 & 0.31 & 0.35 & 0.10 & 6.91 & 1.00 & \ldots & \ldots  & 0.45 & 0.03 & 6.74 & 2.81\\
RWA 24 & 11.01 & 0.28 & 0.24 & 0.29 & 0.07 & 8.82 & 1.92 & $-$3.67 & 0.38 & $-$0.13 & 0.06 & 5.51& 12.02\\
RWA 25 & 10.61 & 0.36 & 0.30 & 0.32 & 0.19 & 3.41 & 0.38 & $-$4.40 & 1.54 & 0.38 & 0.10 & 6.50 & \\
RWA 26 & 13.71 & 1.33 & 1.22 & 1.05 & 0.49 & 6.69 & 1.51 & 7.02 & 1.77 & 1.19 & 0.12 & 19.16 & 5.8\\
RWA 27 & 12.83 & 1.60 & 1.26 & 0.79 & 0.33 & 7.23 & 0.35 & 2.80 & 0.11 & 0.61 & 0.02 & 34.58 & 6.58\\
RWA 28 & 15.08 & 0.17 & 0.09 & 0.08 & 0.17 & 0.96 & 0.06 & $-$0.60 & 0.06 & $-$2.08 & 0.33 & 0.18 & 4.8\\
RWA 29 & 14.26 & 0.16 & 0.20 & 0.23 & 0.10 & \ldots & \ldots & $-$4.00 & 0.88 & 1.14 & 0.21 & 3.61 & LD\\
RWA 30 & 13.78 & 0.18 & 0.10 & 0.11 & 0.16 & 1.02 & 0.06 & $-$1.81 & 0.24 & $-$5.73 & 1.80 & 0.76 & \\
\hline
WISE Stars  \\
\hline
205736.61+522117.0& 10.98 & 0.60 & 0.63 & 0.78 & 0.16 & 37.64 & 29.74 & $-$5.43 & 0.55 & 0.38 & 0.07 & 15.64 & 22/QP\\
205816.27+522832.8& 11.79 & 0.09 & 0.06 & 0.04 & 0.07 & 0.07 & 0.04 & $-$0.01 & 0.02 & $-$0.99 & 0.04 & 0.57 & \\
205842.10+522545.7& 14.53 & 0.82$^b$ & 0.59 & 1.10 & 0.89 & \ldots & \ldots & $-$1.89 & 0.06 & \ldots & \ldots & 11.12 &LD\\
205855.04+523039.6& 15.46 & \ldots & \ldots & 0.40 & \ldots & \ldots & \ldots & \ldots & \ldots & \ldots & \ldots & 2.60 & QP\\
205904.15+523007.8& 13.85 & 0.64 & 0.38 & 0.68 & 0.61 & 0.78 & 0.09 & $-$2.05 & 0.07 &  \ldots & \ldots & 7.50 & LD\\
210021.42+522257.1& 13.88 & \ldots & 0.88 & 1.02 & \ldots & \ldots & \ldots & $-$4.51 & 0.50 & \ldots & \ldots & 52.81 & 39.5\\
210024.05+521451.0& 12.64 & 0.25 & 0.29 & 0.38 & 0.11 & 6.32 & 1.85 & $-$3.21 & 0.30 & 0.83 & 0.07 & 6.16 & QP\\
210035.25+523334.0& 12.37 & 0.04 & 0.03 & 0.05 & 0.07 & 0.81 & 0.05 & $-$0.80 & 0.07 & $-$1.68 & 0.21 & 0.17 & \\
210038.77+522757.5& 15.43 & \ldots & \ldots & 0.56 & \ldots & \ldots & \ldots & \ldots & \ldots & \ldots & \ldots & 2.65 & LD\\
210058.34+522856.1& 10.96 & 1.64 & 1.19 & 0.69 & 0.49 & 3.25 & 0.11 & 1.45 & 0.16 & 0.85 & 0.02 & 27.81 & LD/QP\\
210110.57+521512.9& 11.44 & 0.73 & 0.76 & 0.63 & 0.19 & 8.94 & 1.88 & $-$9.21 & 3.67 & 0.61 & 0.02 & 13.75 & 32\\
210246.43+523114.0& 13.82 & 1.35$^b$ & 0.09 & 0.05 & 1.33 & \ldots & \ldots & $-$0.18 & 0.04 & \ldots & \ldots & 0.02 &\\
210311.54+523124.3& 13.55 & 0.50 & 0.08 & 0.06 & 0.50 & \ldots & \ldots & $-$1.24 & 0.18 & \ldots & \ldots & 0.45 & \\
\hline
Aspin Stars  \\
\hline
CN 3N & 9.68 & 1.14 & 0.72 & 0.56 & 0.43 & 2.62 & 0.06 & $-$1.98 & 0.34 & 0.41 & 0.08 & 17.11& QP\\
Cyg 19 & 9.90 & 0.42 & 0.38 & 0.35 & 0.16 & 3.85 & 0.96 & 3.69 & 1.21 & $-$1.21 & 0.11 & 9.79& 9.5\\
HH381 IRS & 9.72 & 0.15 & \ldots & \ldots & \ldots & \ldots & \ldots & \ldots & \ldots & \ldots & \ldots &\ldots & \\
IRAS 14 & 14.23 & \ldots & 0.64 & 0.34 & \ldots & \ldots & \ldots & $-$2.49 & 0.96 & \ldots & \ldots & 1.80 & QP\\
IRAS 15N & 9.70 & 0.18 & 0.23 & 0.24 & 0.16 & \ldots & \ldots & $-$5.00 & 1.04 & $-$3.19 & 0.93 & 6.79 & \\
IRAS 15S & 11.95 & 0.07 & 0.08 & 0.04 & 0.06 & 2.01 & 0.34 & $-$0.50 & 0.09 & $-$0.89 & 0.12 & 0.67 & \\
\enddata
    \tablenotetext{a}{Period of eclipse}
    \tablenotetext{b}{Dominated by measurement errors}
\end{deluxetable}

\subsection{Types of Variability}

We divide the light curves by taxonomy as a way to gain insight into the physical causes of the changes.   
There are three traits which we focus on: periodicity, changes in color--color space and changes in color--magnitude space. 
Some sources were strongly periodic. Others seemed cyclic, in the sense that their signals when up and down somewhat regularly, but the actual frequency and amplitude did not appear stable. We identified this group as quasi-periodic (``QP'' in  Table~\ref{Table:YSOs}).   Still other sources experienced monotonic changes in one direction for weeks or months, sometimes reversing. We identified these as long duration  (``LD'' in  Table~\ref{Table:YSOs}).   The remainder are considered stochastic.

\begin{figure}
\includegraphics[width=8.in]{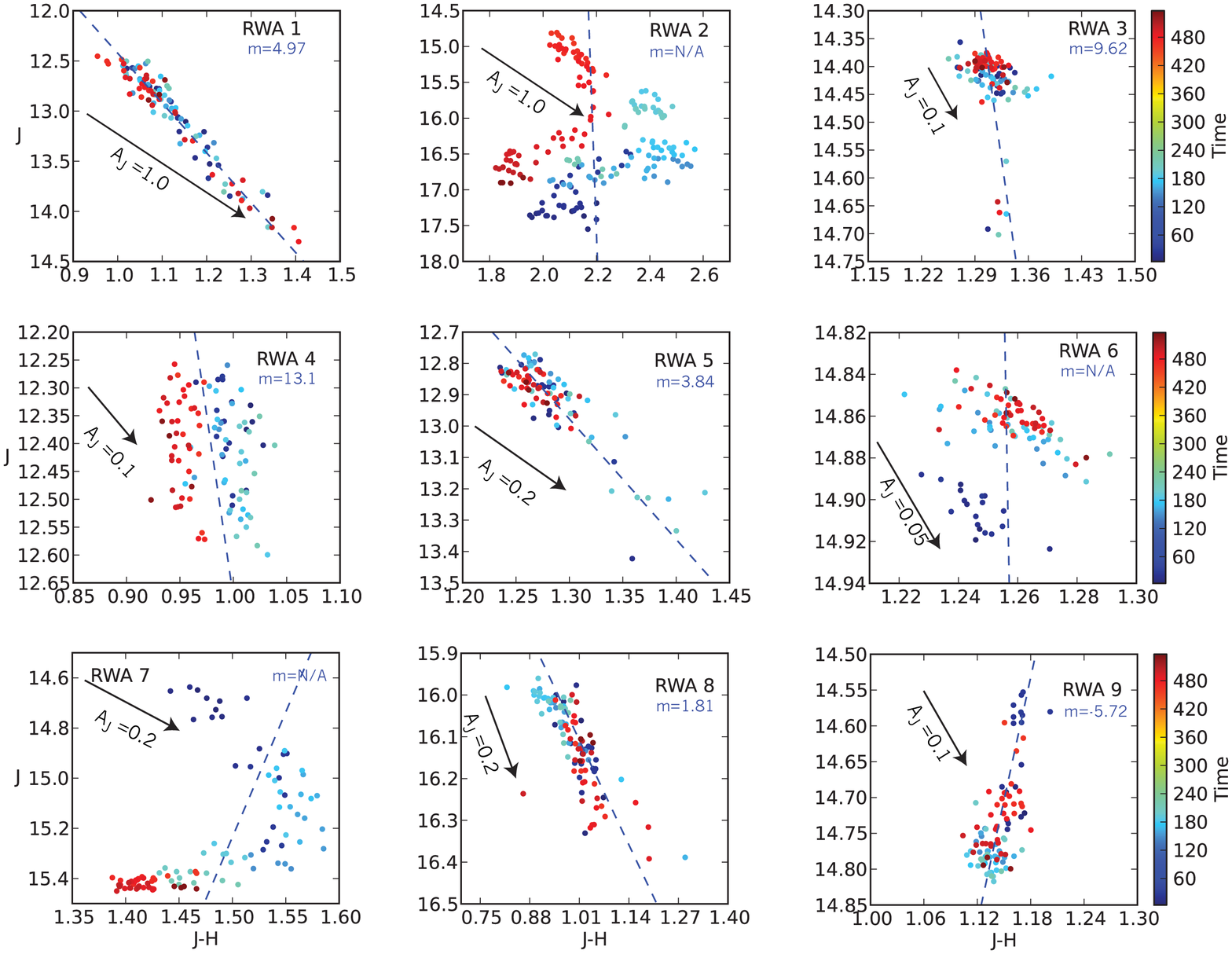}
\caption{Examples of fitted $J$, $J-H$ color--color diagrams for RWA 1--9.  A linear model is not always representative (\eg\ RWA~2 and RWA~7).  The direction and length of the ISM reddening vector is indicated in each window as is the fitted slope (m). Time is in days from April 23, 2008
and is indicated by color.}
  \label{Fig:JJH}
\end{figure}

\begin{figure}
\includegraphics[width=8.in]{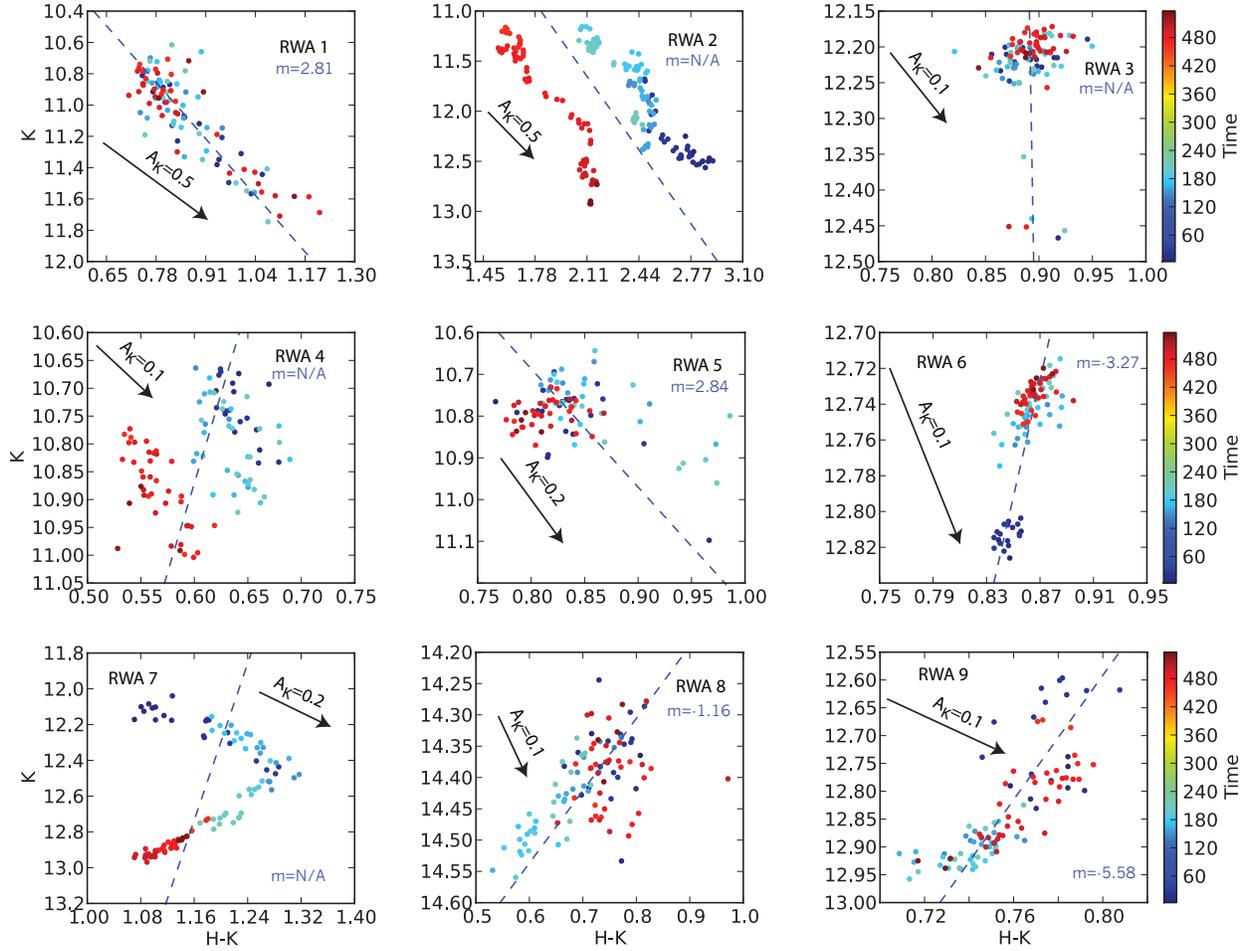}
\caption{Examples of fitted $K$, $H-K$ color--color diagrams for RWA 1--9.  A linear model is not always representative.  The data for RWA~8, RWA~9 and the latter half of RWA~7 have trajectories perpendicular to the reddening vector.  The direction and length of the ISM reddening vector is indicated in each window as is the fitted slope (m). Time is in days from April 23, 2008}
  \label{Fig:KHK}
\end{figure}

In Paper~I we found some sources appeared to follow the CTTS locus in color-color space \citep{Mey97}, while others followed the reddening vectors, still others followed a hybrid track between the two;  finally a few sources moved very little or even had a negative slope.  In the color-magnitude diagrams, most sources moved to the red as sources became fainter, but some moved markedly in the opposite direction (\eg\ RWA~9 and RWA~19; see also Figures~\ref{Fig:JJH} and \ref{Fig:KHK}).  Notably several sources showed {\bf both} behaviors.  We indicate sources that became bluer as they became dimmer with a negative sign on the slopes given in Table~\ref{Table:YSOs}. Below we comment on some of the correlations in these trends.

\begin{figure}
\includegraphics[width=6.5in]{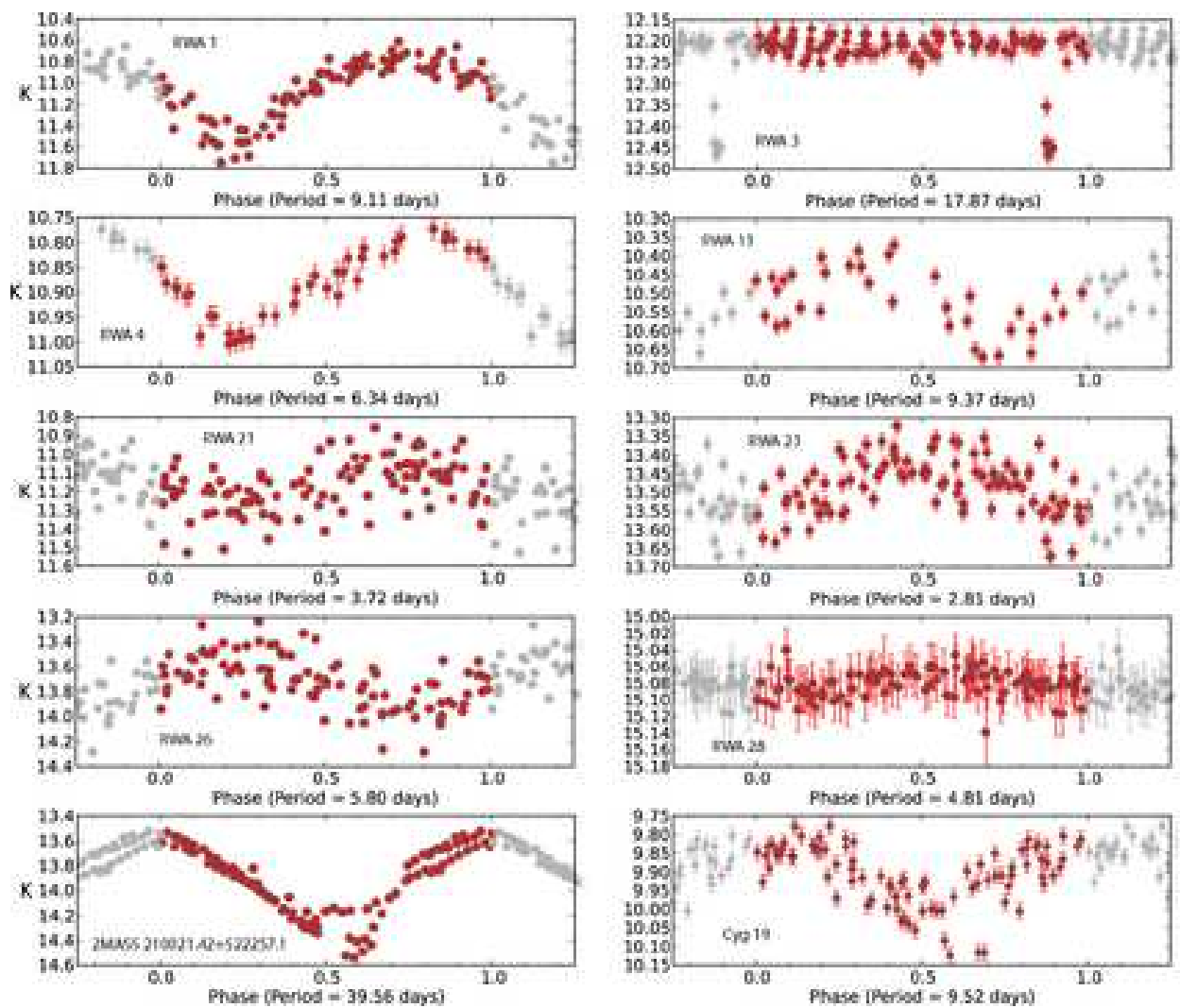}
  \caption{A sample of YSOs with stable periods of  3-40 days.  For each star,  the $K$ band light curve folded over the listed period. The stars are labeled in their individual windows.
   Data for all 3 seasons is used for all stars except RWA~4 and RWA~13 which showed the strongest periods in season three. }
   \label{Fig:Periodic}
\end{figure}

\begin{figure}
\includegraphics[width=6.5in]{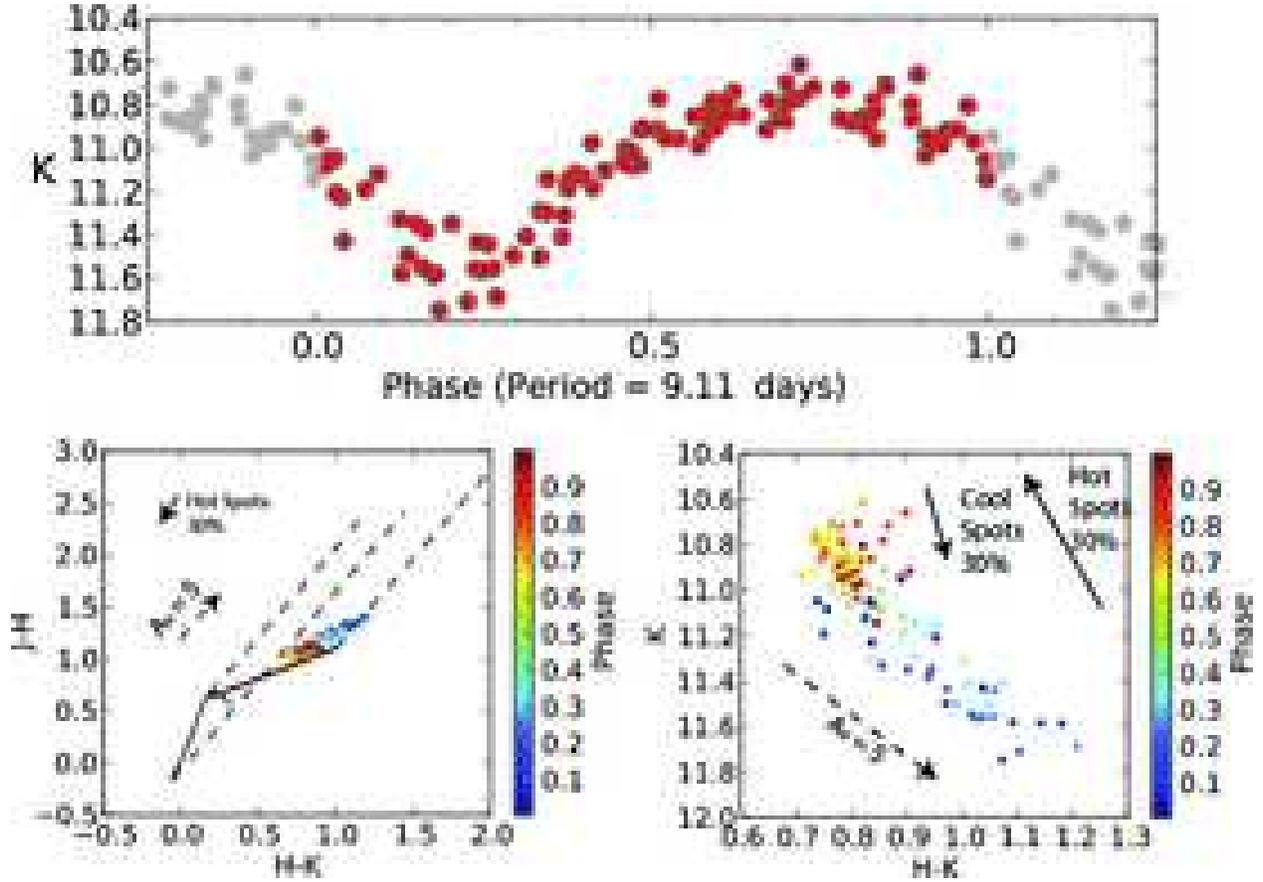}

  \caption{Top: Folded $K$-band lightcurve for RWA~1.
  Bottom-left: color-color diagram for RWA~1. Dashed lines indicate the reddening direction --   5 $A_V$ is indicated. A short arrow indicates the effect of a 30\% filling factor of hot (8000 K) spots.  Cool (2000K) spots have a negligible effect. The solid straight line indicates the ``CTTS locus'' \citep{Mey97}, RWA~1 seems to move parallel to this line.  
  Bottom-right: $K$, $H-K$ color--magnitude diagram for RWA~1.   In addition to 5 $A_V$,  there are vectors indicating the influence of hot  and cool spots with filling factors of 30\% would have on the trajectory in this diagram.
   Phase is indicated in the color version.  
}
   \label{Fig:RWA1}
\end{figure}

\begin{figure}
\includegraphics[width=6.5in]{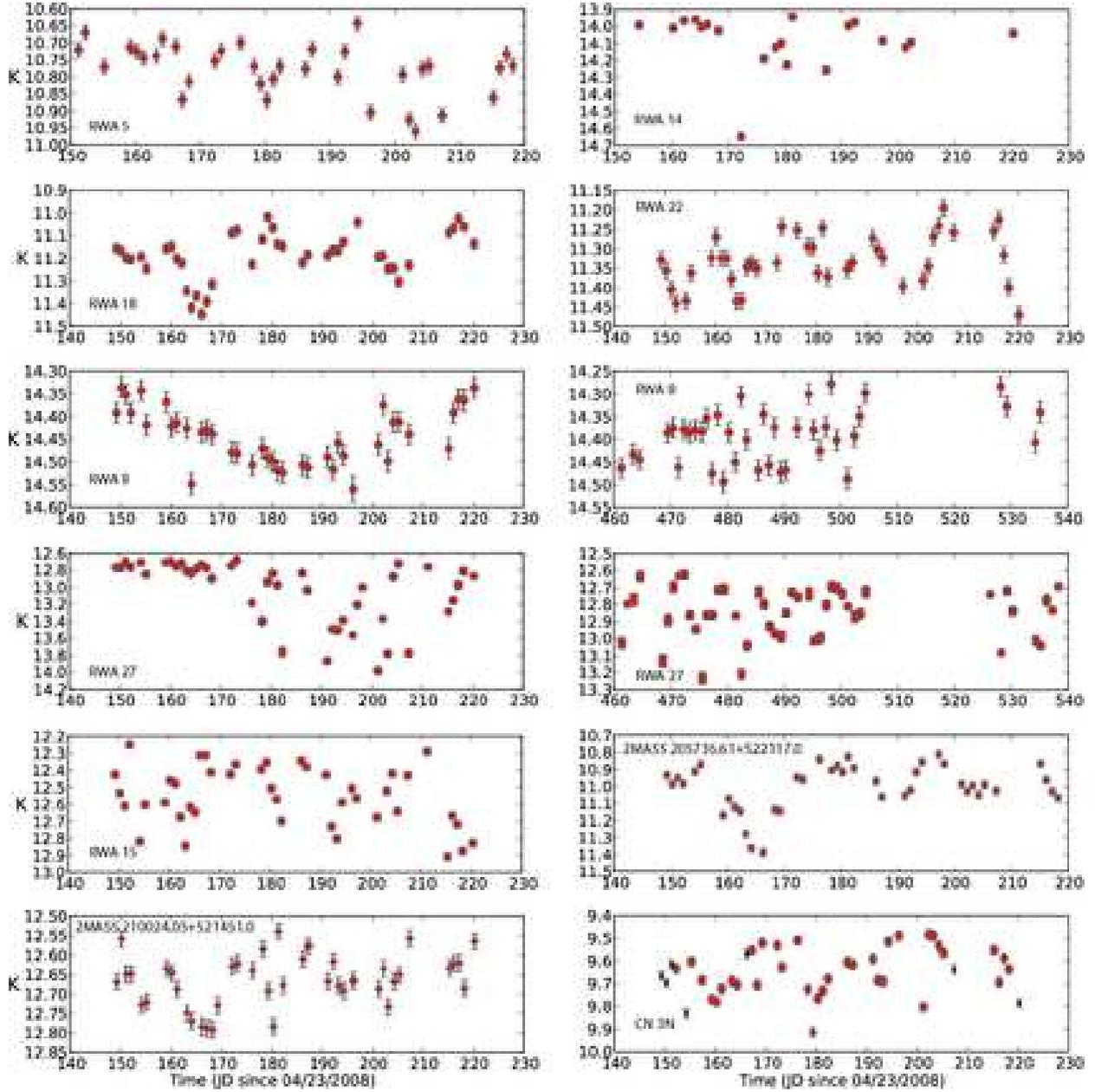}
 \caption{Examples of stars with many up and down cycles but no fixed period. $K$ band data are shown.
 Season 2 lightcurves are shown except for RWA 8 and RWA 27 which have both seasons 2 and 3 shown.
}
   \label{Fig:Quasi-Periodic} 
\end{figure}

\begin{figure}
\includegraphics[width=6.5in]{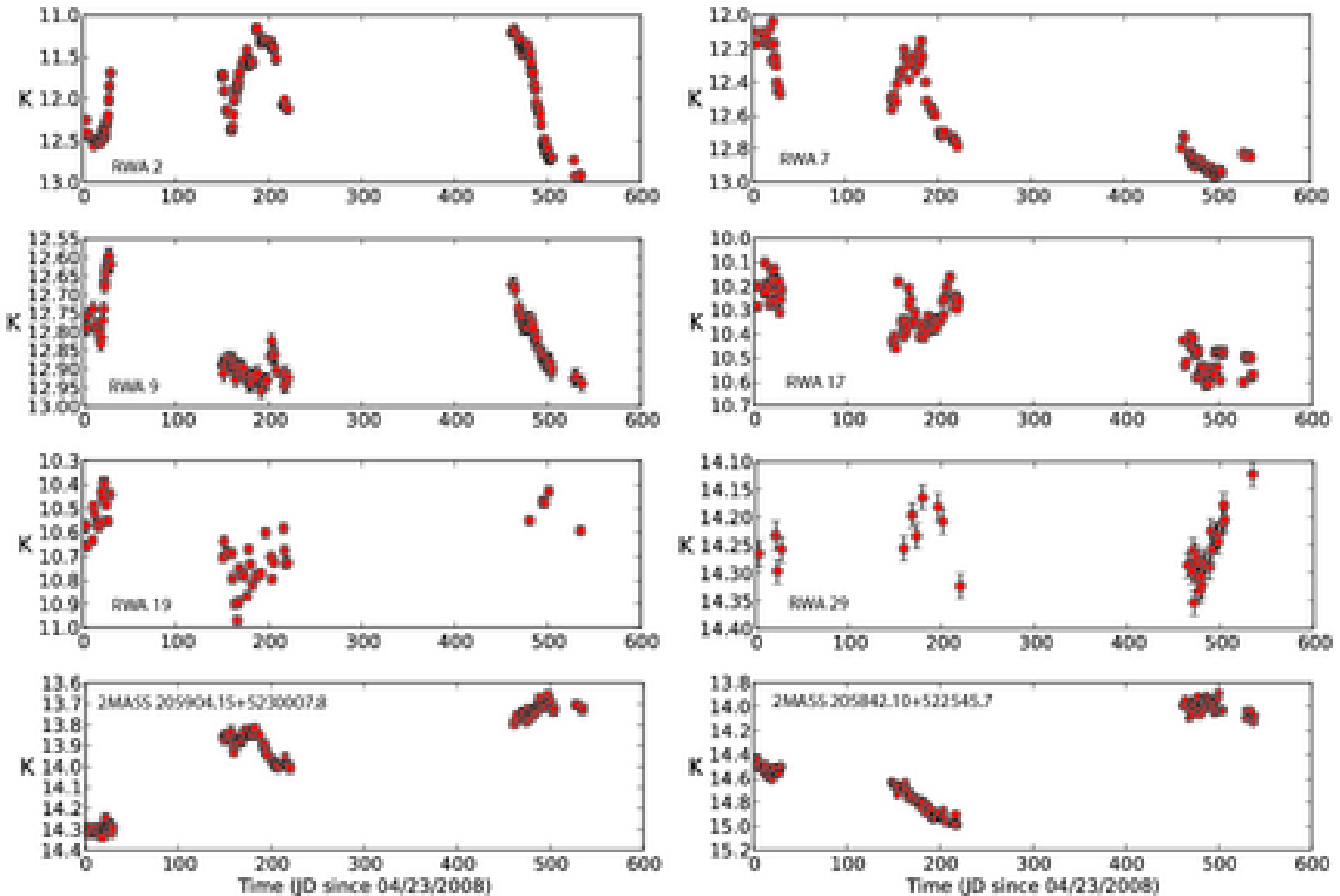}
\caption{Examples of stars with long duration changes. $K$ band data are shown.}
  \label{Fig:LD}
\end{figure}

About one-quarter of the stars are seen to vary with cycles of 2.5 to 10 days (Figure~\ref{Fig:Periodic}). The periodic signal was not as strong as for the eclipsing binaries in the field noted in Paper~II. Typical $K$ band flux changes for these stars were between 0.25 and 1.0 magnitudes, with an RMS deviation off of this signal of 10 to 50\%. 
 These stars became redder as they became fainter.  Often the color changes tracked the reddening vector, but for several stars, the color change tracked the CTTS locus \citep{Mey97}.   RWA~1 is the prototype for this class  (Figure \ref{Fig:RWA1}).  It has a fairly stable 9.1 day period with a $K$-band peak to trough amplitude of about 0.8 mag.  However, this can change by 0.2 mag  cycle to cycle.  The scatter about the nominal trajectory is about 10 times the observational error.
The observed change in the color--magnitude diagram is nearly linear: for 1 magnitude total observed change in $K$, the change in $H-K$ was about 0.5.  The JHK color-color change is highly phase-dependent, with faint phases nearly being consistent with a normal photosphere and bright phases reaching the edges of the CTTS locus. 
 
About  one-fifth of the stars  
appear to be quasi--periodic, meaning they show successive rises and falls in flux that do not occur with a stable frequency or amplitude.
We identify these as  ``QP" in Table~\ref{Table:YSOs}.  Some examples of these are shown in  Figure~\ref{Fig:Quasi-Periodic}.  RWA 8  shows a ``U'' -shaped decline and rise of 0.3 $K$ mag in season~2. 
In season 3, the same star shows a periodicity which ranges from about 4 days to about 10 days and modulates with an amplitude which varies from  0.1 to 0.2 mag.  RWA 15 shows a 19.34 day period, but only in season 2. 
2MASS~210058.34+522856.1 and 2MASS~210024.05+521451.0
(Figure~\ref{Fig:Quasi-Periodic}) show similar behavior with periods of 4.8 and 23.6 days respectively in season 2 and different  periods in evidence in season 3.  2MASS~205855.04+523039.6 
shows a  20.05 day period in season 3 and a 47.29 day period in season 2 -- both with similar periodogram peak power $\sim 12$.  
Interestingly, while most of these stars show relatively small changes in their position in the color-color diagram, 
most of these stars show a tendency to become redder (in $H-K$) as the star becomes {\sl brighter} in the KHK color-magnitude diagram.  We will discuss this further in \S3.2.

Another group of stars were long duration - possibly irregular variables, identified as ``LD" in Table~\ref{Table:YSOs} (see Figure~\ref{Fig:LD}).  These stars could show large changes in $K$ magnitude. RWA~2 showed an amplitude change of 2 $K$ mag and  RWA~7 had a change of 1 $K$ mag. It is not clear that these objects truly form a common group since  they showed a wide variety of trajectories in color space. 
After RWA~2 and RWA~7, other stars with long duration, slow changes  tended to have smaller changes, but several still showed at least half a magnitude change at $K$ including RWA~9,  RWA~12,  RWA~17 and RWA 19.  Four of the newly identified WISE candidates also show long duration high amplitude changes.
Among these, 2MASS~210058.34+522856.1 shows a long term trend of $\Delta K>0.6$ in addition to its short term rises and falls
on times scales which vary from 5 to 15 days.

Some stars had strong variability, but no dominant period. 
For example, RWA~6 was not seen to vary in any one season.  However, it brightened by about 0.1 $K$ mag  between seasons 1 and 2. The net effect was significant correlated variability. IRAS~15N showed a similar one--time shift.  It became fainter by 0.1 mag in season three.  
The amplitude of the changes in  RWA~10 and RWA~11 were only about 0.1 mag RWA~20 and RWA~25 changed by up to 0.4 mag, but these changes were essentially colorless. 
2MASS~205745.36+521331.4 showed a 0.04 mag decline in brightness in season 3, and was about 0.16 mag brighter at $J$ than during the 2MASS survey. 

\begin{figure}
\includegraphics[width=6.5in]{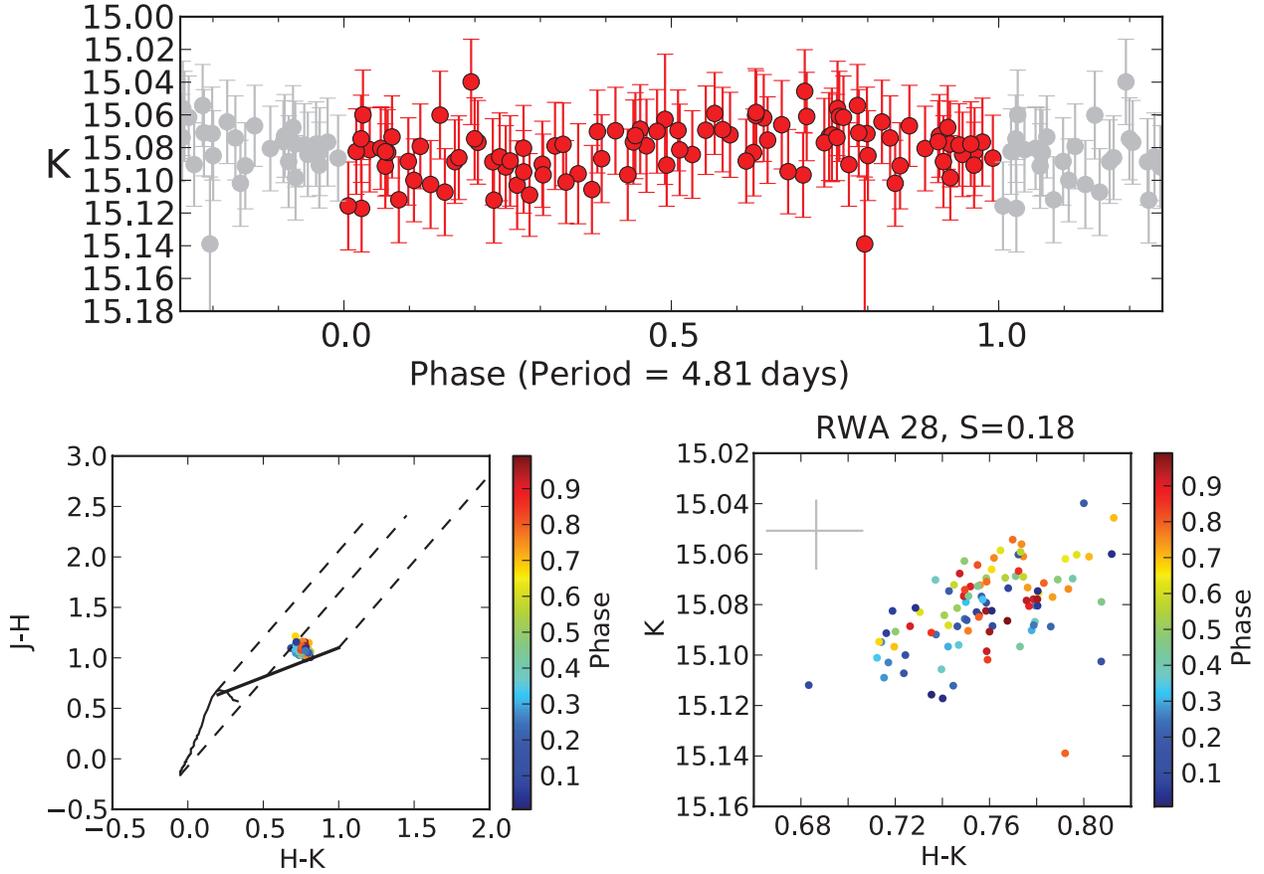}
 \caption{Top: One and a half year's worth of $H-$band data for RWA 28 folded on a 4.81 day period. 
        Left: The small change in the color--color diagram parallels the direction of the K-M portion of the main sequence indicating an effective cooling. 
        Right: The $K$ $H-K$ color magnitude diagram.  The 1$\sigma$ error bar for these data is plotted.  No significant deviation is seen.  Phase is indicated in the color version.  }
     \label{Fig:RWA28}
\end{figure}

A periodogram of RWA~28 shows significant power in $K$ band at a period near 4.8 days. Indeed the data folded on this period show a  signal of about 0.05 $K$ mag which is similar to the errors (Fig.~\ref{Fig:RWA28}). Similar signals are not seen in the other bands -- however these have more noise than $K$ band. Three of the WISE sources were also found to have  $S< 0.55$: 
2MASS~210246.43+523114.0, 2MASS~210311.54+523124.3 and 2MASS~210110.57+521512.9.
The first two of these sources are very faint at $J$ ($\sim$ 19.7 and 18.2 respectively) which works against the identification of weak variability.  The last source varies by less 0.05 mag in any channel.

\subsection{Observed Color Changes}

Strong secular color changes are seen among the YSOs. This is in stark contrast to stars studied in Paper~II,  wherein the minimal color changes were consistent with the eclipse model in which the net effect of the eclipse was to fractionally reduce the contribution from the Wien tail of one of the two stars, or in the case of a pulsation, change the temperature by a few percent.  Neither effect changes the near-IR color significantly.   
The slopes in Table~\ref{Table:YSOs} can be compared to expected values for common cases.
For example,  significant NIR color changes changes would be expected if, for example, the flux change were induced by a portion of the disk blocking the star as in AA Tau. 

Indeed, in the $J$, $J-H$ color magnitude diagram most stars get redder as they get fainter, although the slopes are often somewhat steeper than prescribed by the ISM reddening law  (Figure~\ref{Fig:JJH}).  
This implies less differential reddening or $R > 3.1$ \citep[$R \sim 3.1$ typical of the ISM][]{Rie85}. This result is consistent with studies of massive star forming regions such as $\eta$ Carinae which find R$\sim 4$ presumably due to grain growth \citep{Pov11}.
On the other hand, all show scatter along the extinction track in excess of that expected based on the data quality and more importantly many stars show the opposite trend in the  $K$, $H-K$ color magnitude diagram.


Many stars in the quasi-periodic group grew redder as they grew brighter (as seen in the color--magnitude diagram).  
 2MASS~205736.61+522117.0 follows a regular pattern that becomes about 0.8 brighter at $K$ {\bf and} 0.15 mag redder in $H-K$.
In the color--color diagram the path of the star follows a path between the CTTS locus \citep[$(J-H)/(H-K) \sim 0.55$][]{Mey97}, and the direction of the K-M main sequence ($(J-H)/(H-K) \sim -1.0$). 
The trajectories of several of the stars in the quasi-periodic group parallel the CTTS locus.

Stars in the long duration group showed variations of 0.5 and 1.5 at $K$. The observed color changes were between $\sim 0.1 - 1.5$ in $H-K$.  Stars were seen to get both brighter and fainter as they became redder.  RWA~7 (Figs.~\ref{Fig:JJH} and \ref{Fig:RWA7}) is a source which shows both behaviors. Some of these stars follow the CTTS locus in the color--color diagram, while other more closely parallel the reddening vector or even the K/M branch of the main sequence.   

From an observational perspective, the lightcurves of YSOs are both complex and ordered.    Some are very periodic, but even these have unaccounted noise on top of the periodic pattern.  Others seem periodic briefly, but appear to have other overriding affects. 
All the lightcurves have individual characteristics -- unlike eclipsing or pulsating variables, no two are alike.  On the other hand, when color changes are included, patterns emerge. The periodic  sources typically get redder as they get dimmer, while many of the quasi-periodic sources showed the opposite behavior.  In the next section we investigate possible physical basis for the behaviors.

\section {Discussion}

\subsection {Sources of Variability}
There are many reasons the luminosity of a YSO might change. 
Some mechanisms include extinction changes, cool spots, hot spots, changes in the accretion rate, changes inner disk radius or perhaps even the inclination of the inner disk region.   \citet{Her94} noted that each of these effects would be characterized by different colors and timescales. Changes in the overall extinction may be the simplest to imagine. Perhaps induced by the disk, extinction can cause practically unlimited changes in the apparent flux of a star.  However, such changes should move the star in the direction of the reddening vector. 
For ISM-like dust, an $A_V$ change of 5 mag  corresponds to $\Delta$J and $\Delta$K calculated of about 1.25 and 0.5 respectively with a change of $\Delta (J-H)$ of $\sim$ 0.48 and  $\Delta (H-K)$ of about 0.27 \citep{Rie85,Ind05}.  Motion in this direction clearly occurs in  certain phases of the lightcurves shown in Figures~\ref{Fig:JJH}, \ref{Fig:RWA1}, \ref{Fig:RWA4}, \ref{Fig:RWA2} and \ref{Fig:RWA7}.

Cool spots, like those on the Sun, were first identified as a contributor to the variability of PMS stars in the 1980s \citep{Vrb85}.
They have been used since as a method of measuring stellar periods \citep[e.g.][]{Att92}.  But there is a limit to the variability cool spots can induce, since the spot is typically only  1000-1500K cooler than the nominal photosphere. 
\citet{Car01} examine modulation induced by cool starspots in the NIR.  
They used a fairly simple spot model which assumes a photospheric temperature of 4000 K, appropriate for a 1 Myr, $\sim 0.5 M_\odot$ star \citep {Dan97}.  The change in luminosity at any given wavelength is given roughly as
$\Delta m(\lambda ) = -2.5 log \{1-f [1-B_\lambda (T_{spot}/B_\lambda(T_{*})]\}$
(where B$_\lambda$(T)is the Planck function).\footnote{This starspot model ignores limb darkening, inclination effects, and opacity differences which occur due to the different temperatures between the spot and photospheric temperatures.}
Using a  maximum spot coverage of  30\% the maximum, $\Delta$J and $\Delta$K calculated were about 0.35 to 0.30 respectively with a color change $<$ 0.05. In the I band, the observed luminosity change due to cool spots is typically $<$ 15\% \citep{Coh04}.  
The implied color change due to a lower effective temperature is $<$ 5\%.  Hence in the JHK color spaces, cool spots have almost no effect.  This is consistent with what is seen in some of the least variable objects. 

Hot spots, thought to arise from accretion, can cause a larger signal than cool spots since the temperature difference is typically larger (a factor of 2 or 3 hotter than the surrounding photosphere). 
For warm spots (8000K) and a spot coverage of  30\%, the maximum 
$\Delta$J and $\Delta$K are calculated to be about 1 and 0.6 respectively, with a change of $\Delta (J-H)$ of 0.2 and  $\Delta (H-K)$ of about 0.15 \citep{Car01}. Truly hot spots ($>$10,000K) on K and M stars can induce signals as high at 1 magnitude at $J$ and color changes of 40\% in $J-K$, even with small filling factors \citep{Sch09}.  In the JHK color-color diagram, hot spots and extinction generally move objects in opposite directions, making them hard to disentangle.  

In many cases, the relation of the disk to the accretion appears to be fundamentally important to the observed variability. Changes to a few key individual parameters can explain much of what is observed. \citet{Mey97} examined the positions of dereddened Class II objects located in Taurus  in the JHK color-color diagram.  They noticed that the  dereddened positions corresponded to a line in JHK color--color space --  the cTTs locus.  Furthermore, they showed that the observed colors could be accounted for by a model which combined the stellar photosphere with a variable accretion rate, a variable  location of the inner disk boundary, and variable disk inclination.  
This model is somewhat degenerate in that a single point in the color-color diagram can be achieved with several combinations of the three parameters. While more complex parameterizations of similar models now exist \citep[\eg][]{Rob06}, these only add to the degeneracy.
The degeneracy is broken somewhat by using additional color-space analyses. The $K$, $H-K$ and  $J$, $J-H$ color-magnitude diagrams were shown to be useful \citep[see][esp. Fig 25]{Car01}.

Table~5 
gives the typical observed behaviors expected for the physical phenomena discussed above. 
The values are meant to be representative.  In the first row we indicate the change induced by an increase in extinction.  While we give the values for A$_V$=5,  the full range is effectively limitless. One can envision dust as an acute phenomena wherein a large quantity of dust temporally obscures the star.  Extinction may also be periodic as in the AA~Tau phenomena where in a warp in the disk is regularly or semi-regularly seen to obscure the central star \citep{Bou03, Bou07}. 
 The color vectors, whether $\Delta K$/$\Delta (H-K)$,  $\Delta J$/$\Delta (J-H)$  or  $\Delta (J-H)$/$\Delta (H-K)$, are well known \citep{Rie85, Lad92, Ind05} to within changes due to the particle size distribution in the dust (\ie $R_V$). 
 
In the second row we itemized cold spots; these are well studied.  Over a timespan of a few months,
spots are expected to be periodic with the rotation period of the star, which for young stars is a few days \citep[e.g.][]{Att92, Reb01}. While critical in terms of a visible manifestation of the stellar magnetic field and useful for rotational analysis, cool spots are the weakest and most physically limited of the phenomena listed.
 Hot spots, shown in the third row,  can have more of an impact because the temperature differential can be much higher. The allowed range in high temperature spots may be as high at 12,000K \citep{Her94}. The values in Table~5 
 are given for 30\% filling factors and temperatures of 2000K and 8000K on a 4000K photosphere for the cold and hot spots respectively. Hot spots should also be periodic phenomena as the stars should be brighter on the side with more spots --although there is empirical evidence that hot spots are less stable. 
 
\citet{Mey97} presented separate models of accretion rate, hole size and inclination over a small grid of representative values, the extremes of which are shown in rows 4 and 5.  These effects lead to the star becoming bluer instead of redder as it becomes fainter.  \citet{Bri12} recently modeled deep X-ray observations of TW~Hya and showed that 
hole size and accretion rate may be linked.  They show that to account for some of the observed changes in the NeIX line, the accretion rate needs to have increased by about a factor of 10 {\it and} the hole size needs to have shrunk by a factor of about 2.  The changes occurred over the course of 12 days.  Changing the inclination of a system appears to be an even more efficient way to make the disk become both fainter and bluer \citep{Mey97}.  Of course, it makes no sense for an AU scale disk to precess on a time scale of days or months, however the near-IR is only sensitive to the inner edge of the disk.  A  warped inner edge has the effect of increasing the effective surface area of the disk relative to the line of sight of the observer.  Rows 4 and 5 in Table 5 indicate the expected slopes induced by the changes to the accretion parameters.  A range is listed since the 
two likely change in tandem with the hole becoming smaller as accretion rates increase.

\begin{figure}
  \includegraphics[width=6.5in]{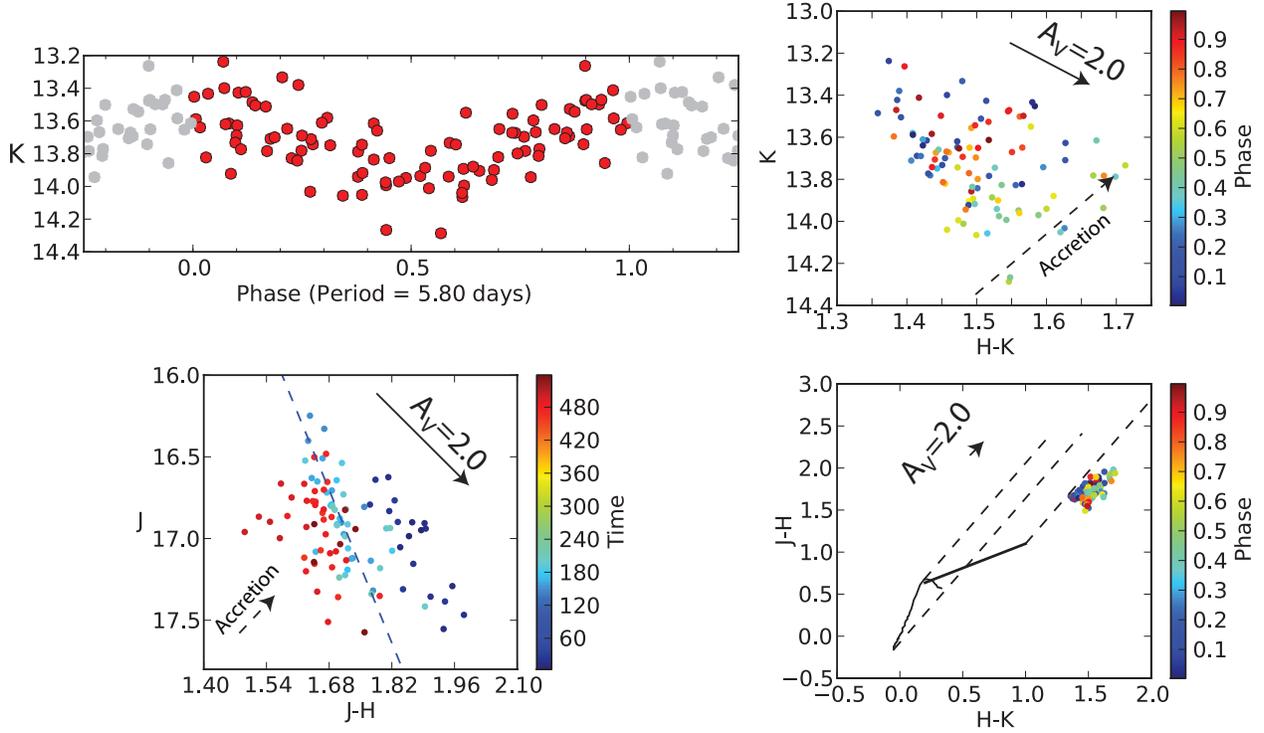}
  \caption{Top-left: K-Band lightcurve for star RWA~26  in which all data from 3 seasons are folded on a 5.8 day period.
  Top-right: The $K,H-K$ color-magnitude diagram for RWA~26 color indicates phase. 
 Bottom-left: The $J,J-H$ color-magnitude diagram for RWA~26 color indicates time. 
 Bottom-right: color-color diagram for RWA~26. Dashed lines indicate the reddening direction while
    color indicates phase. In all color diagrams a reddening vector of 2 $A_V$ is shown.   The dashed line in the upper-right and lower left panel indicates a change in accretion of a factor of about 30 (excluding effects of the change in hole size). 
   The primary mechanism for the change in brightness due to phase of this star seems to be due to reddening as indicated by the upper-right panel, although we note that the reddening appears steeper than predicted.  A long term decrease in the accretion rate is indicated by the bottom-left panel.
}
  \label{Fig:RWA26}
\end{figure}

We can interpret much of the observed variability in terms of these parameters -- spots, extinction and disk changes. The models are not optimal for detailed lightcurve analysis, but they do provide anecdotal evidence of the kinds of behaviors responsible for the observed changes.   In Figures~\ref{Fig:RWA1} and 11 --\ref{Fig:RWA7} we have labeled the lower panels with vectors derived from \citet{Car01}  and \citet{Mey97}.
{RWA~26} (Figure~\ref{Fig:RWA26})  is perhaps the simplest of the group to understand.  The 5.8 day period is very stable and the color shifts in the color--color diagram and the color--magnitude diagram are both dominated by extinction changes.   The slopes in the color--magnitude diagrams are steeper than expected from ISM extinction and the scatter in this signal is still about 20\%.
The data shown in the lower-left panel of  Figure~\ref{Fig:RWA26}  shows that RWA~26 moves nearly along the reddening vector, while the 
lower-right panel of Figure~\ref{Fig:RWA26} shows a similar effect -- an increase in extinction by $\sim$ 5.0 $A_V$\footnote{In these figures we use the reddening law from \citet{Rie85} with $R_V=3.1$ and $A_K = A_V* 0.112$.}  at the intermediate phases compared to phases 0 and 1.
There is considerable spread in the data by nearly 30\% above what would be expected if extinction was the only source of the changes. This indicates other factors at work. We further notice that the overall data distribution moves up and to the left in time in both color magnitude diagrams, slowly over the 550 days of observing (most clearly in the lower left panel where color indicates time, not phase).  This is indicative of a steady increase in the accretion rate by, perhaps, a factor of 30 \citep{Car01}.  

The data shown in the lower-left panel of Figure~\ref{Fig:RWA1} show that RWA~1 moves nearly along the cTTs locus.  This is different from RWA~26 and indicates that RWA~1 has a similar extinction change but less of a change in the disk and accretion characteristics.  The 
lower-right panel of Figure~\ref{Fig:RWA1}  shows that phases 0.0-0.4  are consistent with a drop in extinction by $\sim$ 5 A$_V$.
This is followed by a pattern that is consistent with a periodic decrease in cool spots and an increase in hot spots.  
So it appears that the extinction is changing in a way that is nearly cancelled out by the change in accretion in a 9.11 day period. 
But this is not the only plausible explanation for the observed changes. 
Even with this additional interpretation, the noise on the signal is $\sim20\%$ which is much higher than expected based on the observational statistics, and is indicative of additional factors at work.

The case of RWA~4  (Figure~\ref{Fig:RWA4}) also can explain some of the noise.   The star is clearly periodic with a period of about 6.35 days. Extinction changes by 2.0 in A$_V$ as is shown in the color-color plot in the figure, but extinction appears independent of phase.  On the other hand, the lower--right panel shows extinction change to be part of the periodic cycle.   Higher extinction is seen between phase 0.1 and 0.4. The remainder of the data can be understood by asserting that the size of the hole in the inner disk changes by a factor of two  -- out of phase with the extinction.  Such a hole would induce $\pm 10\% $ noise on the extinction based signal.


\begin{deluxetable}{p{4.7cm}rrccccr}
\tablewidth{0pt}
\tablecaption{Expected observed properties of different physical changes.} \label{tab:props}
\tablehead{
\colhead{Type of variability }
  &\colhead{$\Delta$ J}
    &\colhead{$\Delta$ K}
      &\colhead{Typical Period}
      &\colhead{$\Delta J     \over {\Delta (J-H)}$}
      &\colhead{$\Delta K     \over {\Delta (H-K)}$}
      &\colhead{$\Delta (J-H)	 \over {\Delta (H-K)}$}
      &\colhead{Refs.}
      }
\startdata
Extinction ($\Delta A_V =5$)      &    1.25   &          0.5    &       non-periodic$^a$   &   2.6   & 1.8 &    1.7 &  1\\
Cold spots &                  0.15  &          0.1   &             4--12 days  &     $\sim 5$  &      $\sim 11$  &   $\sim 1.5$      & 2,3\\
Hot spots     &               1.5           & 0.6 &            4--12 days        &    $\sim 2$  &      $\sim 5.7$  &   $\sim 1.5$            & 2,4 \\ 
$^b$Change in accretion rate  log $-$8.5 -- log $-$7 $M_\odot$ yr$^{-1}$  &   0.6 & 0.75 &   ?  &  $-4 :-5 $     &   $-2:-5$ & $0.4:0.6$  &     1,2 \\
$^b$Change to inner disk edge  radius 1--4 R$_*$  &      0.5 &    0.75 &     ?  &       $-1.25: -5$      &   $-1.4: -4$ & $\sim 0.35$  &     5 \\
\enddata
\tablenotetext{a}{Changes in extinction could be imagined both as long term trends which would not be periodic, or as features in a disk which could be periodic or quasi--periodic.}
\tablenotetext{b}{The resultant slope in color space due to changes of accretion rate depend on the inner hole size and the inclination of the inner disk. Likewise,  the slope of a trajectory due to changes inner hole size depends on the accretion rate and the inclination of the inner disk.}
\tablenotetext{~}{References: 1- \citet{Rie85}, 2- \citet{Car01}, 3- \citet{Sch05}, 4- \citet{Sch09}, 5- \citet{Mey97} 
}
\end{deluxetable}

\subsection {Objects with Similar Light Curves}
With these physical concepts in mind we revisit the observational results from \S3. 
 We start with the strongly periodic stars similar to RWA~1 (Figure~\ref{Fig:RWA1}).  In the color--magnitude diagrams, the color changes appear to be dominated by reddening with some evidence of both warm and cool spots.  In the color--color diagram, the object moves parallel to the cTTs locus.  \citet{Mey97} argue this is indicative of changes either to the accretion rate, hole size or inclination.  However, the result that $K$-band flux is at a minimum when the star is at the extreme end of the cTTs implies a maximum disk cross-section between the star and the Earth when accretion is high.  This is reminiscent of the  AA~Tau phenomenon in which the rotation of a circumstellar disk with a high latitude ÒwarpÓ can periodically obscure the star
(Bertout 2000; Bouvier et al. 2003). Morales-Calder\'on \e (2011) remind us that any process which creates overdense asymmetric regions in the inner disk could produce the flux dips.  Stars can also be obscured by ÒcloudsÓ of relatively high opacity in the disk atmosphere or geometric warps in the inner disk. 

\begin{figure}
\includegraphics[width=6.5in]{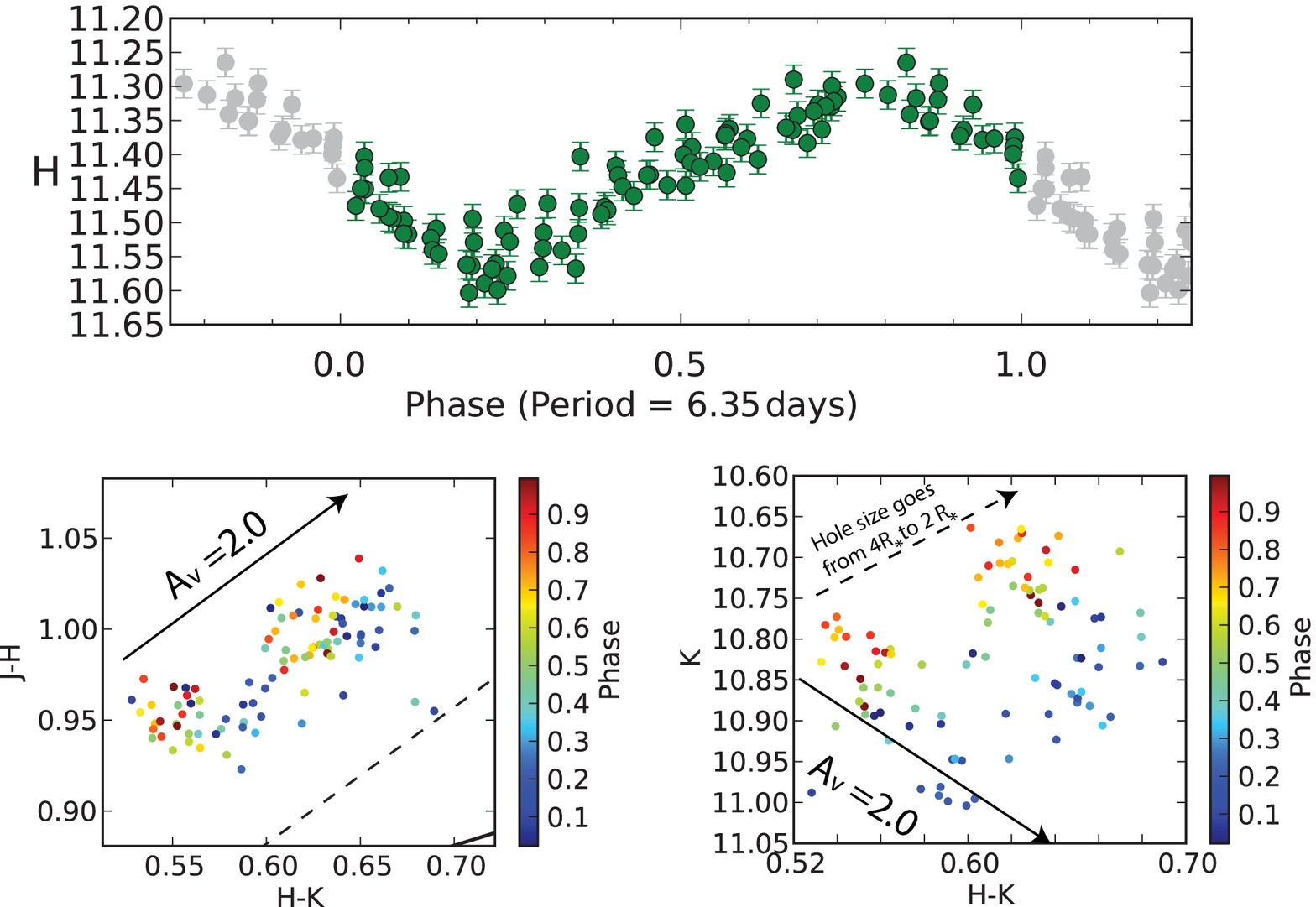}
 \caption{Top: One and a half year's worth of $H-$band data for RWA 4 folded on a 6.35 day period. 
                 Left: The $J-H$, $H-K$ color-color diagram  shows a change equivalent to $A_V=2.0$ but this does not seem phase dependent.
                 Right: $K$, $H-K$ color magnitude diagram shows a circular pattern followed by the star.  Arrows indicating the direction influence of changes in extinction and hole size are given for guidance.    Phase is indicated in the color version.
                  As indicated in Figure~5, the star gets bluer (by 0.1 at $J-H$) for a given magnitude over the 500 days of observation.}
     \label{Fig:RWA4}
\end{figure}

We class RWA 1 with other stars that show a dominant period and a trend toward reddening as the K-band flux drops.
While there are at least 7 other such sources, no two are alike in terms of period, amplitude or color range. On the other hand, a couple are worth mentioning in particular. 
The K band magnitude of {RWA~4} dims by $\sim$0.3 mag over the 1.5 year observations while$H$ dims by about 0.1 mag and J does not noticeably change.  At the same time, it has periodic variability of 0.2 magnitudes in K (0.35 in J) with a period of about 6.34 days (Figure~\ref{Fig:RWA4}). The long-term and periodic components of this variability are orthogonal in color-mag space. 
The flux of the star shows a constant downward trend with a mean $K$ of 10.75 in season 1 and 10.9 in season 3.  While the $\sim$ weekly color cycles indicate a change in extinction, the long term trends indicate a drop in accretion.

\begin{figure}
\includegraphics[width=6.5in]{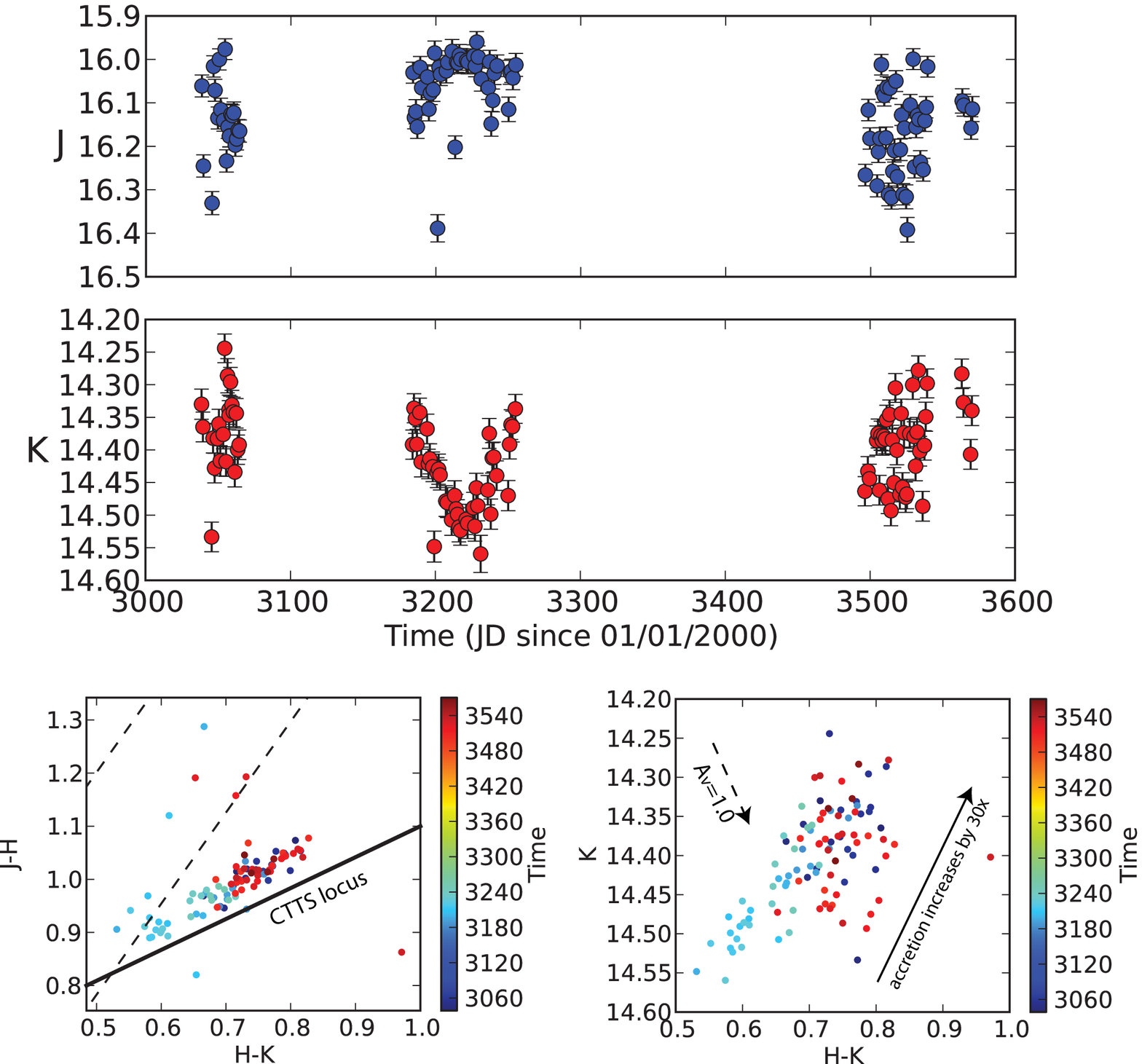}
  \caption{Top: One and a half year's worth of $J$ and $K$-band data for RWA 8.
                 Left: The $J-H$, $H-K$ color-color diagram  shows a change parallel to the CTTS locus. 
                 Right: $K$ $H-K$ color magnitude diagram shows the bulk of the change can be accounted for by asserting an increase in the accretion rate. A small increase in extinction is also needed to account for deviations from the model.  The $J$, $J-H$ data shown in Figure ~5 imply an A$_V$ change of about 1 (A$_K \sim 0.1$).
Time is indicated in the color version. }
                      \label{Fig:RWA8}
\end{figure}

Unlike the periodic sources, many of the quasi-periodic sources were noted to become redder as they become brighter in the $K$, $H-K$ color-magnitude diagram.  RWA~8 (Figs. 5 and \ref{Fig:RWA8})  is typical of this group.  Rapid cyclic variations are seen, sometimes on timescales of a few days and sometimes on timescales of months.  There is a general trend in the color data which shows the star moving along the cTTs locus and taking on the appearance of an increasingly active cTTs.  The  $K$, $H-K$  color-magnitude diagram shows that the behavior is consistent with an increase in the accretion rate of a factor of about  30 with a concurrent increase in the extinction of 0.1-0.2 A$_K$. The $J$, $J-H$ changes are almost exclusively reddening,  at first decreasing by $\sim 0.7$ A$_V$ and then increasing by $\sim 1$~A$_V$. Other stars, including  RWA 15, RWA 20, RWA 22 and RWA 24 also showed similar behavior of weak, erratic  quasi-periods and quasi-phased motion in the color--magnitude diagram. 
  For these stars $0.1 < \Delta (H-K) < 0.3$ and $0.1 < \Delta K < 0.7$.
  The data are  consistent with accretion rate changes $\sim 30x$ with extinction changes of a few of A$_V$.
Another star in this group  is an eclipsing binary, RWA~3, which has a 17.87 day period. The 2\% eclipsing system rate among the YSOs in the region is consistent with the  $\sim 1\%$ rate seen in the field. No disked source was found to have a period $<2$ days.  
The eclipsing binary nature of this star may make spectroscopic probes of this behavior possible.

Most of the observed signals are not periodic at all and need to be understood in terms of unique events or long period trends.
Among the most fascinating stars were two of the long duration objects -- RWA~2 and RWA~7 (Figs. \ref{Fig:RWA2} \& \ref{Fig:RWA7}). RWA~2 exhibits changes of 3 mag at $J$ and full period cycles on time scales of 2 months.  The color data appear continuous between seasons 1 and 2, but follow a different track in season three.   One explanation of these data is a steady decrease in dust obscuration by about 1.0 $K$ mag, followed by a sudden drop in the accretion rate and then a slow return to the original level of extinction.  Incorporating 2MASS data, we find 
the total range in J is at least 4 mag.  The $J-K$ color varies from 2.9 (2MASS) to about 5 when the star is at its faintest. The trajectory of the star in JHK color-color and color magnitude space is indicative of changes in both reddening (diagonal lines in the color--magnitude diagram) and accretion rate (the distance from the lower-left corner in the color--magnitude diagram or the distance from the reddening band in the color--color diagram).  In $J$,$J-H$ space the data from the first 2 seasons follows a steep negative slope ($\sim -2$), data in the third season have a similar slope, displaced by about 1 J mag brighter and starts with a brief period which resembles decreasing extinction.   $L^\prime$ imaging and $K$ band spectroscopy show RWA~2 has strong H$_2$ and Br$\gamma$ emission and is the member of a resolved binary (700 AU [0.9"] separation) with a very red companion ({\it C.A. Aspin unpublished}).   The observed mid and far IR fluxes of RWA~2 are consistent at the 10-20\% level from the IRAS epoch until the current one. Thus, most of the flux variations are confined to the near-IR. We fitted SED models 
 to the median JHK results in the near-IR as well as Akari, IRAS  Herschel data at 12, 18, 25, 60, 70, 90, 140, 160, 250, 350 and 500\micron.
   From the
best fitted model of the spectral energy distribution of RWA~2 using the models of  \citet{Rob06} we derive an age for the RWA~2 protostar of $< 100,000$ yr,  A$_V \sim 15$, an envelope mass of order a solar mass, which is similar to, or greater than, the mass of the central object.  

\skipthis{
\begin{figure}
\includegraphics[]{/data/swolk/SCIENCE/props/CHANDRA/AO14/RWA2/ModelRWA2.eps}
 \caption{The best model fit of the spectral energy distribution of RWA using the models of \citet{Rob06}.  The total flux is in black and fitted to the median JHK results in the near-IR as well as Akari, IRAS  Herschel data at 12, 18, 25, 60, 70 90 140 160, 250, 350 and 500\micron.  The purple curve is flux due to accretion, green - the disk, red - the envelope while the dashed line is the unreddened  photospheric flux.}
 \label{Fig:SED}
\end{figure}
}

\begin{figure}
\includegraphics[width=6.5in]{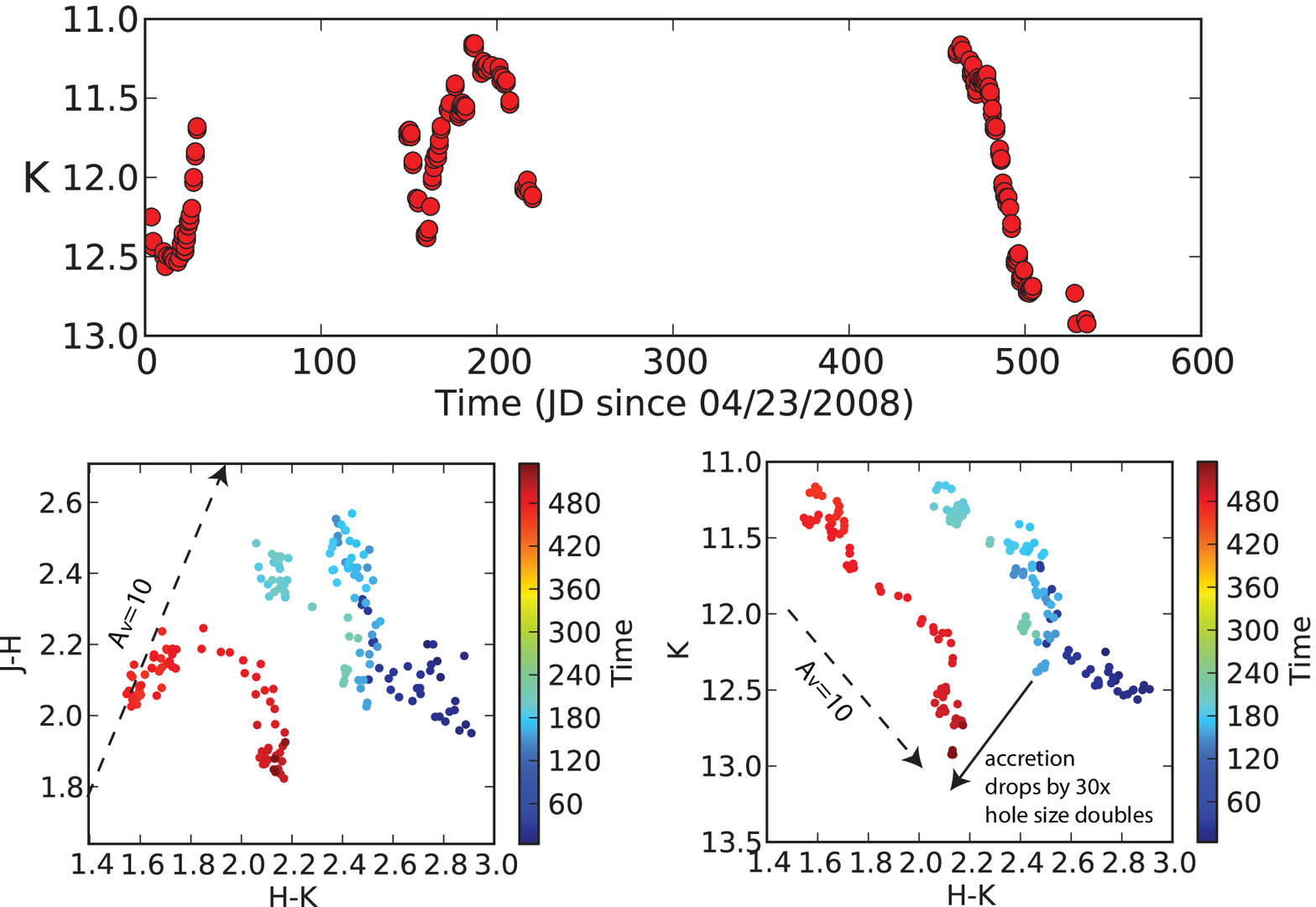}
 \caption{Top: One and a half year's worth of $K$-band data for RWA 2.
     Left: The $J-H$, $H-K$ color-color diagram  shows changes which neither follow the reddening vector nor the CTTS locus. 
                 Right: $K$, $H-K$ color magnitude diagram shows the bulk of the change can be accounted for by asserting a decrease  the reddening by  $A_K=1$, followed by a dramatic change in the accretion and disk parameters between seasons 2 and three and then an increase  the reddening by  $A_K=1$. 
Time is indicated in the color version. }
     \label{Fig:RWA2}
\end{figure}

RWA~7 is similar to RWA~2 in that the overall observed change was about 1 $K$ mag.  Further, the observed lightcurve changes did not appear monotonic.  The data from seasons one and two show local maxima (Fig.~\ref{Fig:RWA7}).   The color-color diagram, on the other hand, indicates a steadier process -- an apparent circulation.  The color starts near 1.1, 1.45 ($H-K$, $J-H$)  moves fairly steadily toward 1.30, 1.55. down to 1.2, 1.45, arriving back near  1.1, 1.45.  However, despite having the same colors at the end as the beginning the star is clearly not in the same state. The $K$, $H-K$ color-magnitude diagram gives the clearest insight. In this panel of Fig.~\ref{Fig:RWA7}, we see that the trajectory of the star is dominated in the first season by a change in extinction of about 4 A$_V$.  This trend continues through  about Oct 15, 2008 when the trajectory suddenly shifts - coincident with a small double peak in the light curve.  After this time,  and for at least the next year, the trajectory follows a very steady path which can be described as a decrease in the accretion rate by a factor of  $\sim$30 and a doubling of the size of the central hole. 	 In $J$, $J-H$, the trajectory is similar, though not as sharp. The uninterrupted nature of the color--magnitude diagrams hints at a periodic phenomena in which we have missed a near integer number of half-cycles.  For example, we may have missed a $\sim 100$ day minimum  between seasons 1 and 2 and a full $\sim 200$ day period between seasons 2 and 3.

\begin{figure}
\includegraphics[width=6.5in]{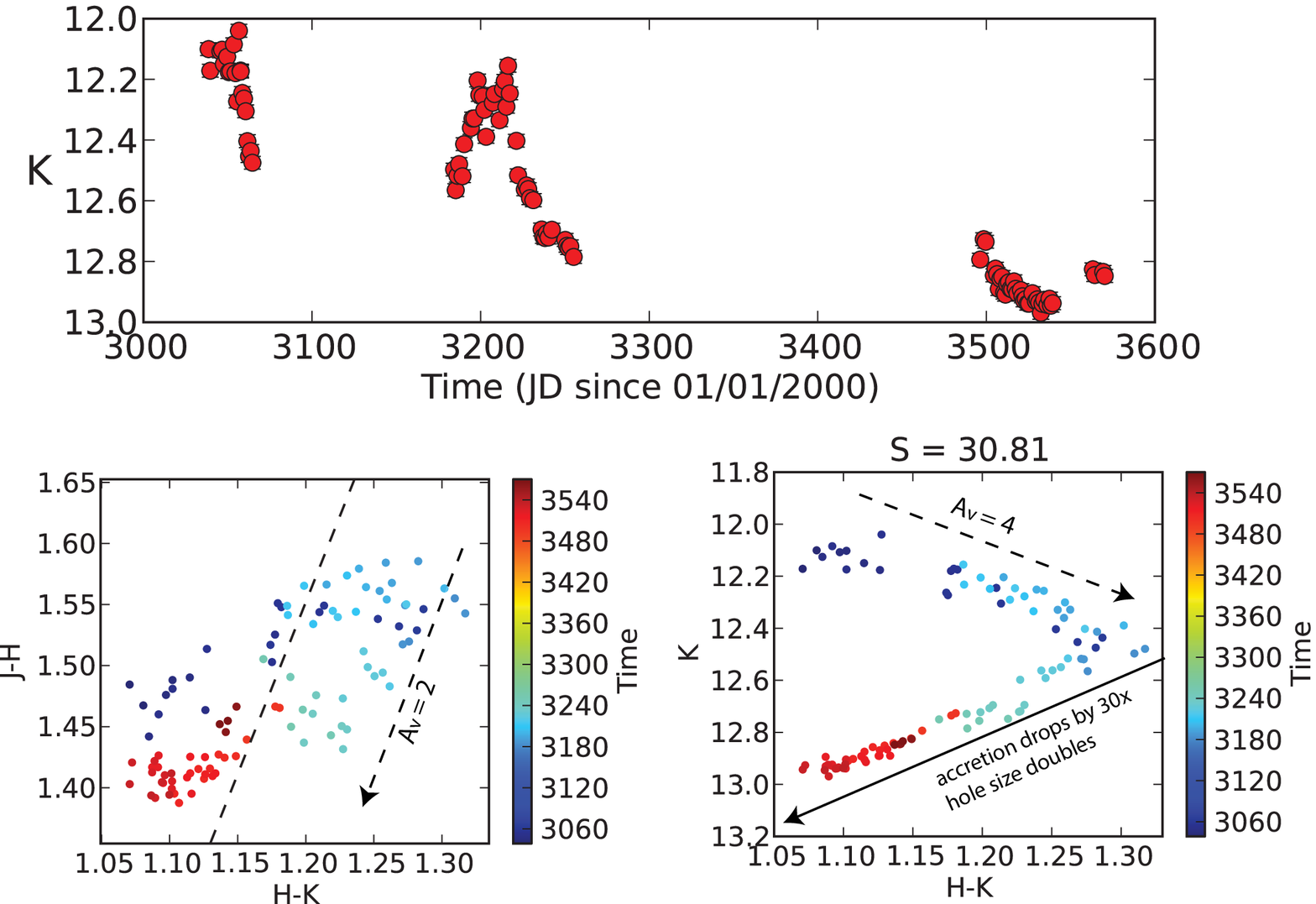}
 \caption{Top: Lightcurve showing one and a half years worth of $K$-band data for RWA 7.
Bottom-Left: The $J-H$, $H-K$ color-color diagram. 
                 Bottom-Right: $K$ $H-K$ color magnitude diagram shows  the change can be accounted for by asserting a decrease  the reddening by  $A_K=0.4$, followed by a dramatic change in the accretion and disk parameters.
Time is indicated in the color version. }
     \label{Fig:RWA7}
\end{figure}

\subsection {Unusual Objects}
One of the difficulties in studying variability in the NIR is that the origin of variability can be very complex and the different effects can either amplify or dampen the combined impact on the resultant lightcurve depending on circumstances.  As a result, there are many lightcurves which appear interesting  but are not easily explained.  Probably the most unusual was RWA~27 (Fig.~\ref{Fig:Quasi-Periodic}).  The star was reasonably well behaved with a $K$ magnitude of about 12.9$\pm 0.3$ and $H-K$ = 0.65 $\pm 0.1$ for all but about 50 days of the observed epochs.
However, starting about October 15, 2008 the star suddenly showed overnight changes of up to 1 magnitude at $K$.  Further,  $\Delta(H-K)  >1.05$
during this period.  The median $J$ mag, which was typically about 14.6$\pm 0.1$,  fell to 16.4.  Overall this is fairly consistent with a compact dust--like occulting object.  This object would be optically thin, with a maximum extinction of about 1 A$_K$  that intersected our line of sight every few days obscuring increasing fractions of the star and then passing out of our line of sight again.   RWA~19 showed a similar, but less extreme behavior: it took 4 days to drop 0.3 $K$ magnitudes with $H-K$ color changes consistent with reddening,  and then recovered just as quickly. In this case the $J$ color change was very similar to the $H$ band change and hence not consistent with ISM reddening.

\subsection {The Scope of the Cluster}
Of the 13 WISE YSO candidates monitored, 9 (69$\pm23$\%) were found to be variable in the near IR bands, via the $S$ index.  This is approximately consistent with the rate of variability found in the Paper~I and \citet{Asp09} samples.   This gives us strong confidence in the WISE selections outlined in \S2.1, including candidates listed in Table~4 which are outside the monitored field. The sources outside the field are found equally to the west and east of the RWA field.  In Table~4 we identify the groups of multiple YSOs within an arcminute of each other. The ``Western group'' is a clump of 6 candidates near 20:26:52 +52:20, three of these are identified as Class I sources.  The ``RWA~2 tip'' is a clump of 4 candidates at the end of the same portion of Lynds~1004 which contains the remarkable object RWA~2.  Three of these objects are identified as Class I.  A final group ``Southclump'' is comprised of three stars within 10\arcsec\ of 21:03:38+52:17:40 containing a Class I and two Class IIR stars.  The compactness of this Southclump raises concerns about the independence of the photometry at the longer wavelengths. 

\citet{Mye12} has recently produced a model of protostar mass and luminosity evolution in clusters which can be used to estimate cluster age. The model assumes a constant protostar birthrate and core-clump accretion.  Model parameters reproduce the initial mass function and match the protostar luminosity distributions in nearby star-forming regions.  Figure 9 in their paper presents a calibration of the model to  measure cluster ages and from the observed numbers of protostars and Class~II objects.  In Cyg OB~7 we have found  32 clear Class II sources and 30 clear Class I sources with the remainder showing colors somewhat between the two.   Using this model in an ``intermediate case"
and dividing the Class~IIR sources evenly between protostars and Class II objects we obtain a cluster age of between 1.0 and 0.5 Myr for Cyg~OB7.   

\section{Conclusions}
	\label{sec:conclusion}

In this paper,  we have studied the NIR variability of young stars in the Cyg~OB7 cluster based on $\sim$ 100 nights of monitoring spaced over about 550 days.   We started by enhancing our candidate list using the recently released WISE point source catalog as well as a previously published catalog from \citet{Asp09}.  In total, we identified 96 total YSO Candidates of which 66 were in the monitored field. The number of these objects which were of the appropriate brightness for monitoring was 49.   Of the 49, 41 were found to have strongly correlated variability in the monitored bands via the Stetson index. There were three additional cases of marginally correlated variability.   If we include the marginal cases, the result is consistent with entire YSO population being variable in the near-IR on timescales of $<$ 10 years, usually on timescales of $<$ two years.  This compares to less than 2\% of the field stars being variable over the $\sim$ 2 year epoch.  We have discovered one eclipsing binary YSO in the field, RWA~3.  It is a good candidate for followup spectroscopy to determine the mass and evolutionary state of the individual members. 

We then studied the shape of the light curves and noticed some emergent patterns.  First, a little over one-fourth of the sources appeared periodic with periods of a few days.  In general, these periodic sources showed color trends consistent with reddening.  The color changes tended to be a bit steep in the sense that there was more change in brightness for a given color change than expected for ISM reddening.  This implies less differential reddening than seen in the ISM (\ie R $> 3.1$).  Another subset of sources identified as ``quasi-periodic,'' showed patterns of brightening and then becoming fainter with cycles of a few days, but the duration and amplitude of the brightening differed cycle to cycle.  Several of these sources showed color changes perpendicular in the color-magnitude diagram to that expected by reddening.   A third group of sources showed large changes of $>$ 0.5 $K$ mag. These were generally smoothly varying with times scales over a month.   A plurality of sources showed purely stochastic variations.  

We argue that changes in a limited range of parameters -- disk obscuration, accretion rate, hole size, inner disk inclination -- could account for much of the variability that was seen.  In the case of the periodic sources, it appears the obscuration of the star by a non-uniformly thick disk was the primary cause of the changes in brightness, but other factors were changing as well.  In the quasi-periodic cases,  the trajectory in the color-magnitude diagram which was perpendicular to reddening implies changes to the accretion rate and/or to the disk structure.  Meanwhile, the long term changes show some periods dominated by slow changes in the disk--accretion structure.  It is clear that multiple processes are taking place in some sources.  Nowhere is this clearer that RWA~7, in which the trajectory in the color magnitude diagram has an inflection point.   While more detailed modeling in the future could be used to provide a better understand of each of the various type of variability, it is equally clear that differences in the state of each star and its disk system are sufficient that each YSO needs to be considered as a unique entity.
 
Finally, we briefly discussed the distribution of YSOs in the region combining the 96 sources in the samples from Paper~I, \citet{Asp09} and the WISE candidates added here. We find that the WISE candidates within the monitoring field are co-located with the previously known YSOs, primarily in the L 1003-1004 clouds.  The sources outside the field  continue that, extending the active star formation to the extreme eastern end of Lynds 1004 and the extreme western end of Lynds 1003.  Empirical modeling indicates the whole star forming region is under 1 Myr in age, with at least one source, the highly absorbed and wildly variable RWA~2, is fitted with an age $<$ 0.1 Myr.

\section{Acknowledgements}
S.J.W. is supported by NASA contract NAS8-03060 (Chandra).  T.S.R.  was supported by Grant \#1348190 from the Spitzer Science Center.
Thanks also to the NSF REU program for funding part of this research via NSF REU site grant \#0757887.
This publication makes use of data products from the Wide-field Infrared Survey Explorer, which is a joint project of the University of California, Los Angeles, and the Jet Propulsion Laboratory/California Institute of Technology, funded by the National Aeronautics and Space Administration. This research has made use of the NASA/ IPAC Infrared Science Archive, which is operated by the Jet Propulsion Laboratory, California Institute of Technology, under contract with the National Aeronautics and Space Administration.  The authors wish to recognize and acknowledge the very significant cultural role and reverence that the summit of Mauna Kea has always had within the indigenous Hawaiian community. We are most fortunate to have the opportunity to conduct observations from this sacred mountain.

\end{document}